\documentclass[prl,preprint,showpacs,byrevtex,endfloats]{revtex4}

\usepackage{epsfig}
\usepackage{graphicx}
\usepackage{dcolumn}
\usepackage{bm}
\usepackage{amssymb}
\usepackage{amsmath}
\usepackage{amstext}
\usepackage{longtable}
\usepackage{amsmath}

\newcommand{\be}{\begin{equation}}
\newcommand{\ee}{\end{equation}}
\newcommand{\ba}{\begin{eqnarray}}
\newcommand{\ea}{\end{eqnarray}}
\newcommand{\baa}{\begin{eqnarray*}}
\newcommand{\eaa}{\end{eqnarray*}}

\begin{document}
\def\BibTeX{\rm B{\sc ib}\TeX}

\preprint{simon et al.}

\title{
Fourier Transform Scanning Tunneling Spectroscopy: the possibility
to obtain constant energy maps and band dispersion using a local
measurement
\\}

\author{L. Simon$^{1}$\footnote[1]{
laurent.simon@uha.fr}, C.
Bena$^{2,3}$\footnote[2]{cristinabena@gmail.com}, F. Vonau$^{1}$,
M. Cranney$^{1}$ and D. Aubel$^{1}$}
\affiliation{$^{1}$Institut de Science des Mat\'{e}riaux de Mulhouse IS2M-LRC 7228\\
4, rue des Freres Lumi\`{e}re, 68093 Mulhouse CEDEX, France}
\affiliation{$^{2}$Institut de Physique Th\'{e}orique, CEA/Saclay,
Orme des Merisiers, 91190 Gif-sur-Yvette CEDEX, France}
\affiliation{$^{3}$Laboratoire de Physique des Solides, Universit\'{e}
Paris-Sud, 91405 Orsay CEDEX, France}

\date{\today}
\begin{abstract}

We present here an overview of the Fourier Transform Scanning
Tunneling spectroscopy technique (FT-STS). This technique allows
one to probe the electronic properties of a two-dimensional system
by analyzing the standing waves formed in the vicinity of defects.
We review both the experimental and theoretical aspects of this
approach, basing our analysis on some of our previous results, as
well as on other results described in the literature. We explain
how the topology of the constant energy maps can be deduced from
the FT of dI/dV map images which exhibit standing waves patterns.
We show that not only the position of the features observed in the
FT maps, but also their shape can be explained using different
theoretical models of different levels of approximation. Thus,
starting with the classical and well known expression of the
Lindhard susceptibility which describes the screening of electron
in a free electron gas, we show that from the momentum dependence
of the susceptibility we can deduce the topology of the constant
energy maps in a joint density of states approximation (JDOS). We
describe how some of the specific features predicted by the JDOS
are (or are not) observed experimentally in the FT maps. The role
of the phase factors which are neglected in the rough JDOS
approximation is described using the stationary phase conditions.
We present also the technique of the T-matrix approximation, which
takes into account accurately these phase factors. This technique
has been successfully applied to normal metals, as well as to
systems with more complicated constant energy contours. We present
results recently obtained on graphene systems which demonstrate
the power of this technique, and the usefulness of local
measurements for determining the band structure, the map of the
Fermi energy and the constant-energy maps.

\end{abstract}


\maketitle

\section{Introduction}
One of the most remarkable feats achieved with an STM, beside the
possibility to visualize material surfaces with atomic resolution,
was the possibility to image the standing waves associated to the
interference of quasi-free electron wavefunctions. This has been
achieved for the first time on a copper surface, for electrons
confined in a circular resonator created with iron ad-atoms,
structure well known as a ``quantum coral''
\cite{Crommie93,Eigler94,Heller94,Manoharan2000}. This observation
has provided direct evidence that electrons are associated to
waves, thus demonstrating the wave-particle duality which is one
of the fundamental concepts of quantum mechanics.

The ``standing waves'' arising in the presence of surface
inhomogeneities are also known as Friedel oscillations
\cite{friedel58}, where the term of Friedel oscillations has been
first introduced to describe the asymptotic dependence of the
perturbed density of states of a free electron gas in the presence
of disorder. Their observation allows the illustration of some
very important concepts of condensed matter physics.  Thus, the
analysis of Friedel oscillations provides a direct observation of
screening and of electron-electron interaction. Moreover, these
oscillations lie at the foundation of the description of the
indirect coupling between magnetic moments via the conduction
electrons in a metal with the famous Ruderman-Kittel-Kasuya-Yosida
(RKKY) interaction potential \cite{RudermanetKittelPR54,
YosidaPR57, Kasuya56}, as well as of the long range adsorbate
interaction mediated by a two-dimensional electron gas
\cite{KnorrPRB02}.

Following their first description by Friedel, the possibility to
use the Friedel oscillations to probe the electronic structure of
materials was considered by many others, notably in the case of
transitional metals \cite{GautierPR65}. They were mentioned in
relation to magnetic impurities in bulk (3D) materials, for which
an important damping factor of the amplitude of the oscillations
is observed:  the oscillations fall off with the distance $r$ as
$1/r^{\alpha}$ where $\alpha$ is the dimensionality of the
considered electron gas. Furthermore, the observation of the
Friedel oscillations was for the first time done indirectly by the
observation of the coupling between two magnetic layers with a
non-magnetic spacer (host metal) \cite{Petroff91} for which the
theoretical description \cite{Bruno91} has revealed the Fermi
surface of the spacer.
 The development of the scanning tunneling
microscopy (STM) has offered the possibility to study the standing
waves in the local density of states (LDOS) which are in fact
\textit{energy resolved Friedel oscillations} for 2D or 1D
electron gases which provide longer coherence lengths than those
observed in 3D. The local density of surface states has been first
obtained by Hasegawa and Avouris \cite{HasegawaPRL93} on confined
Shockley states of Au(111) surfaces at room temperature, and by
Crommie et al. \cite{CrommieNature93} on Cu(111) at 4K.

Subsequently, the dispersion relation $E(k)$ of surface-state
electrons of Ag(111) and Cu(111) permitted the estimation of the
surface state inelastic lifetime \cite{JeandupeuxPRB99}. As large
energies are accessible both below and above the Fermi level, it
was also possible to study the deviation of the free-electron-like
parabolic dispersion, moreover B\"{u}rgi et al. \cite{BurgiPRL02}
have defined a method to directly image the potential landscape on
Au(111) by STM.

For these measurements, the standing waves have been imaged not only at
the Fermi level but also at different energies. This requires the spatial mapping of
$dI/dV$ at different applied voltages $V$, which allows one to focus on individual values
 of the energy specified by $V$, and avoids an integration over all the wavelengths corresponding
  to the energies between the considered energy level and the Fermi level. This
was clearly demonstrated by Schneider et
al. \cite{dieterSchneider2003} on Ag(111). In this study, standing
waves are generated by step edges, and one focuses on a
particular reciprocal lattice direction when analyzing the Fermi surface.

In 1997 Sprunger et al. \cite{SprungerScience97} demonstrated the
possibility to use STM to directly image the Fermi surface contour
of a metal surface, by performing the power spectrum (Fourier transform (FT)) of a
topographic image of a complex ``electron sea'' pattern generated by
a random distribution of point-like surface defects. These studies
were performed around the Fermi level (low bias voltage) and on
simple isotropic Fermi surface of Au(111) and Cu(111) surfaces
\cite{PetersenJESandRelMat00}. In this seminal paper Friedel
oscillations appear in the Fourier transform as a circular feature
centered around the Brillouin zone center with a radius
corresponding to $2k_{f}$. This corresponds to the simple case of
an isotropic Fermi contour centered around the center of the
Brillouin zone. As we will see in the next section, this is a
direct evidence of the singularity in the Lindhard susceptibility
function of the two dimensional electron at scattering momentum
vector $\overrightarrow{q}=2\overrightarrow{k}_{f}$.

The dependence of the Lindhard susceptibility on
$\overrightarrow{q}$ was subsequently elegantly demonstrated by
the observation of standing waves on Be (10$\overline{1}$0)
surfaces \cite{HoffmannPRL97}. In this system the Fermi surface is
no longer isotropic and standing waves can not be generated by all
step edges (i.e. in all directions). As we will see later this
corresponds to the fact that the wavevectors for the
quasiparticles with a given energy can isotropically point in any
direction for a circular Fermi contour centered around the
$\Gamma$, but acquire a dependence on the direction and position
in momentum space if the Fermi surface is no longer isotropic, or
if it becomes split. 

More recently the FT-STM technique (obtaining and analyzing the Fourier transform of STM images) has begun to be applied also to LDOS images ($dI/dV$ maps) which combine STM imaging and a spectroscopic measurement of the LDOS as a function of energy and position. This technique has been denoted FT-STS:  "FT-Scanning Tunneling Spectroscopy", though in the literature the terms FT-STM and FT-STS are sometimes both used to describe FT-STS measurements. The FT-STS technique has been applied on high-$T_c$ superconductors, for a small range of energy around the Fermi level (see e.g.\cite{HoffmanScience02,McElroyNature03,Vershinin2004}). Another
important observation made using the FT-STS was about the spin of
the quasiparticles. As two waves with opposite spin directions
cannot generate quasiparticle interferences (QPI), Pascual et al.
\cite{PascualPRL04} demonstrated the ability to probe indirectly
the orientation of the spin associated to a Fermi surface of a
half magnetic material. We have tested the strength of this
technique on a semi-metal $ErSi_{2}$, for which some constant
energy contours (CEC's) are split into several bands, and we have
demonstrated as well the possibility to determine the whole 2D
band structure in a wide range of energy \cite{VonauPRBR04,
VonauPRL05}. We have shown that the power spectrum features can be
easily explained on the basis of a joint density of states (JDOS)
approach, by a simple geometrical construction formally
established in \cite{simonJCondMat07}.

Here we provide a background for the FT-STS technique by reviewing
some previous theoretical and experimental results. Thus, besides
the JDOS approach, we describe also how the phase factors can be
taken into account in describing the FT-STS features. We underline
that the stationary phase conditions lead to some selection rules
in the scattering events, as considered in the case of the Kohn
anomalies developed for example by Roth et al.
\cite{RothPhysRev66} for the theoretical determination of RKKY
interactions for non-spherical Fermi surfaces. A more accurate and
complete theoretical T-matrix calculation is also presented. We
show then how this technique applies to epitaxial graphene on SiC.
Furthermore, the determination of the band structure and of the
Fermi velocity of graphene quasiparticles, the possibility to
identify the position of the impurities, as well as the
possibility to predict a large extension of the Van Hove
singularity in epitaxial graphene with intercalated gold clusters
are discussed.

\section{Experimental technique}
Our experiments were performed with a LT-STM from Omicron at 77 K
at a base pressure in the $10^{-11}$ mbar range. The $dI/dV$
images were acquired using a lock-in amplifier and a modulation
voltage of $\pm 20 mV$. The graphene samples were prepared in UHV
by the annealing of n-doped SiC(0001) at 900 K for several hours
and subsequent annealing at 1500 K \cite{VanBommelSurfSci75,
ForbeauxPRB98,SimonPRB99,berger}. This preparation method leads to
the formation of a buffer graphene layer covalently bonded with
the substrate and a monolayer graphene decoupled from the
substrate \cite{LaufferPRB08}. The epitaxial graphene has an
intrinsic n-type character and the Dirac point is at $0.4$ eV
below the Fermi level \cite{OhtaSci06, BergerSci06, PremlalAPL09}.
The deposition of gold on graphene was carried out at room
temperature using a homemade Knudsen cell calibrated using a
Quartz Crystal Microbalance. The sample was further annealed at
1000 K for 5 min \cite{PremlalAPL09}.

\section{FT-STS measurements on Au(111) and $ErSi_2$ and their interpretation by the joint-density-of-states approximation}

\subsection{Background}
The underlying principle of the FT-STS technique stems from the
screening of electrons around a localized impurity.  As described
in textbooks \cite{Ashcroft,ziman}, when a positive charge is held
in a free electron gas, this charge will attract electrons,
creating a surplus of charge which screens its electric field. It
is common to solve the Poisson equation for this system in the
momentum space. The response of the system can be described by the
dielectric constant, which at a given momentum $\vec{q}$ is
reduced to $\varepsilon (\vec q) = 1 - \frac{{4\pi \chi (\vec
q)}}{{\varepsilon _0 q^2 }}$, where $q=|\vec{q}|$ and the $\chi
(\overrightarrow{q})$ susceptibility is
$\overrightarrow{q}$-dependent in the reciprocal space. The
susceptibility can be calculated using first order perturbation
theory. In this approximation we consider
 that all eigenstates are "mixed" in the scattering process. Thus the Lindhard theory applies, and the susceptibility is given
by \cite{Ashcroft}:
\begin{equation}\label{1} \chi (\vec{q}) =
\sum\limits_{\vec{k}} {\frac{{f(k) - f(|\vec{k} + \vec{q}|)}}{{E_{\vec{k}}-E_{\vec{k}+\vec{q}}}}}
\end{equation}
where $f(k)$ is the Fermi-Dirac distribution function. The
summation is performed over all possible $\vec{k}$ vectors
providing that the state is occupied, as enforced by the
Fermi-Dirac distribution. In the limit of vanishing $q$ one
obtains the Thomas-Fermi approximation, while for larger values of
$q$, comparable to the value of the Fermi momentum at $T=0K$, the
susceptibility can be calculated explicitly, the result depending
on the dimensionality of the electron gas. For a two-dimensional
electron gas the susceptibility is given by:
\begin{equation}\label{2} \ \chi ^{2D}
(\vec{q}) = n^{2D} (E_F )\left[ {\sqrt {1 - \left( {\frac{q}{{2k_F }}}
\right)^2 } \theta (q - 2k_F )} \right] \end{equation}
where
$n^{2D} (E_{F})$ is the two dimensional density of states at the
Fermi level. The susceptibility is no longer analytic for
$q=2k_{F}$.
Here we have considered a
nearly-free electron gas ($E(\vec{k})=\hbar^{2}k^{2}/2m^{*}$), for which the energy-resolved electron density oscillates with a wave-vector length
$\overrightarrow{q} = 2\overrightarrow{k}$.

\subsection{Measurements of the FT-STS of the LDOS in Au(111) and $ErSi_2$}

\begin{figure}
\begin{center}
\epsfig{file=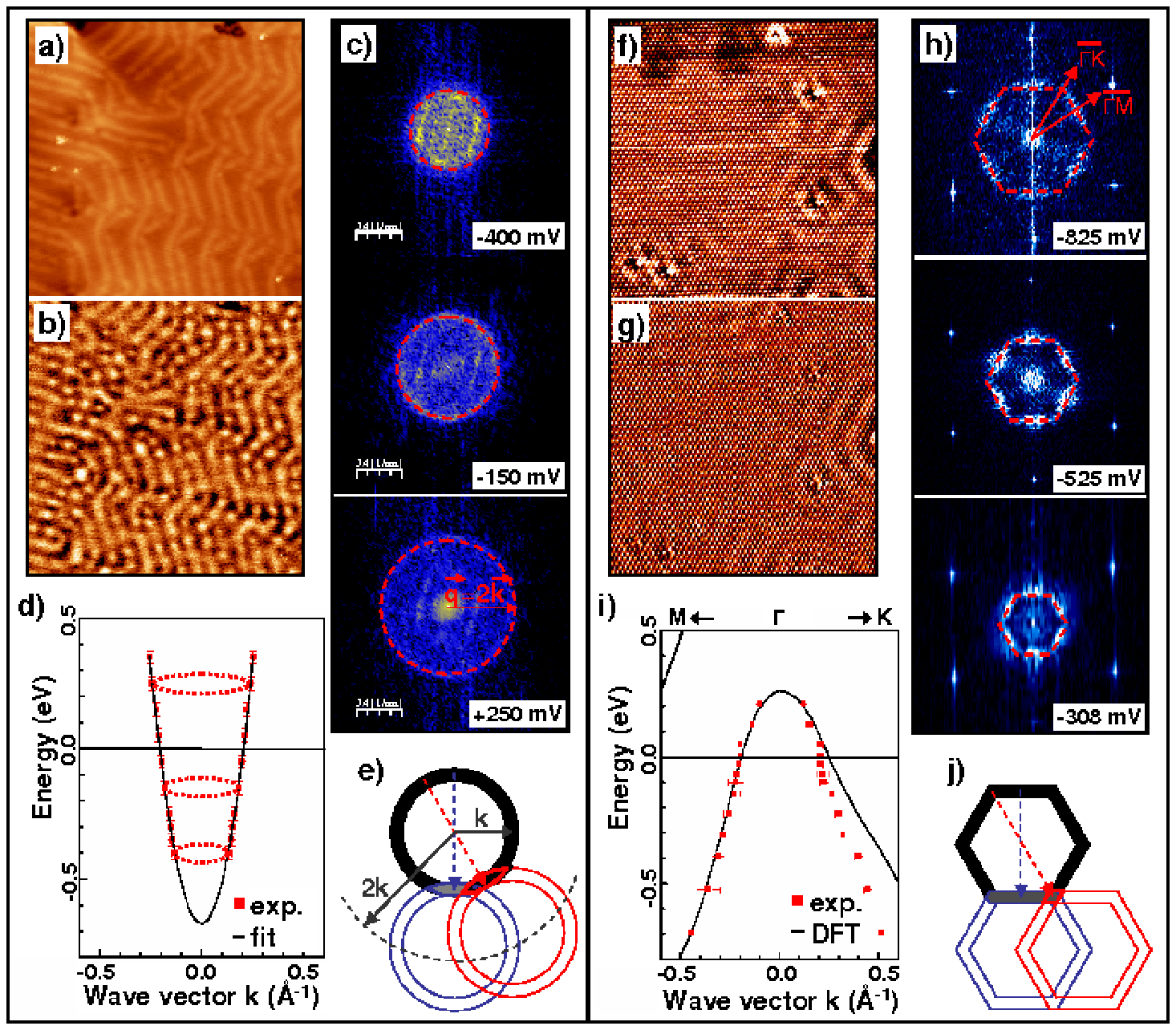,width=8cm,angle=0}
\end{center}
\caption{The STM picture a) shows the herringbone reconstruction
of the Au(111) surface island, perturbed by a nearby dislocation
($50 \times 50 nm^{�}$, -0.4 V). b) is the dI/dV map of the
topographic image a). Three 2D-FFTs of these dI/dV maps (with the
corresponding applied bias) are shown in c), from which we get the
energy dispersion curve of Au(111) shown in d). f) Topographic and
g) dI/dV map at -30 meV ($20 \times 20 nm^{�}$) showing both
standing waves and atomic resolution on $ErSi_{2}$ islands on
Si(111). Three 2D-FFTs of these dI/dV maps (with the corresponding
applied bias) are shown in h), from which we get the energy
dispersion curve of $ErSi_{2}$/Si(111) shown in i). e) and j)
represent the geometrical construction to get access to the
density of scattering events of a CEC for gold and erbium
disilicide respectively.  }\label{Fig1}
\end{figure}

A first experimental illustration of the FT-STS technique is described in figure \ref{Fig1} for
 Schockley states probed on Au(111) surface. In a) the
topographic STM image shows the herringbone reconstruction, and
many defects (scatterers) are visible on the surface. The dI/dV
map taken with a lock-in amplifier allows to measure directly the
energy-resolved local density of states at several energies. We do
not limit our analysis at the Fermi energy, but we analyze several
different energies. When we perform the Fourier transform (here
the power spectrum) of the different dI/dV images at increasing
bias voltage we observe (in c)) a circle which increases in size.
The radius of the circle corresponds to the
$\mid\overrightarrow{q}\mid=2\mid\overrightarrow{k}\mid$ as shown
in the geometrical construction e). We report the value of $q$ as
a function of energy, which yields directly the dispersion curve,
which is here parabolic, as expected in the case of a quasi "free"
electron gas. This is a direct proof of the susceptibility
singularity at momentum
$\mid\overrightarrow{q}\mid=2\mid\overrightarrow{k}\mid$.

This construction can be generalized for non circular CECs. In the
right part of figure 1, for comparison we show the FT-STS for a
two dimensional electron gas of $ErSi_{2}$. The system
investigated here is the Erbium disilicide grown on Si(111)-7x7.
This system has been extensively studied theoretically and
experimentally by photoemission \cite{StaufferPRB93,
WetzelSSCom92}, and an atomic model of this structure has been
determined by Gewinner and co-workers \cite{TuilierPRB94,
StaufferPRB93}. This structural model consists in a p(1x1) plane
of Erbium atoms inserted between the silicon substrate and a
buckled silicon layer \cite{WetzelSSCom92}. This system leads to a
perfect 2D metal-semiconductor interface which shows a
semi-metallic character with a hole pocket centered around the
center of the surface Brillouin zone $\overline{\Gamma}$ and six
electron pockets around the $\overline{M}$ symmetry points at the
Fermi level. As shown in figure \ref{Fig1}.i) the expected
dispersion is "hole-like" because it shows a reversed parabolic
dispersion indicating a "negative" effective mass. Indeed,
contrary to the case of Au(111) surface, the size of the feature
in the FT decreases when approaching the Fermi level.

In f) and g) we show typical topographic and dI/dV map images with
point defects and standing waves pattern. Here the waves are non
longer circular but have a hexagonal shape. In h) the power
spectrum shows clearly a hexagon whose size decreases when
approaching the Fermi level. The direct consequence of this CEC
topology and of the fact that the susceptibility is
$\overrightarrow{q}$ dependent, is the reinforcement of intensity
along specific directions. Indeed along $\overline{\Gamma M}$ the
FT is brighter than along $\overline{\Gamma K}$. While at first
glance it may seem that this could be an effect of the form of the
matrix elements in the scattering process, it can however be
explained entirely by the topology in the CECs. As schematized in
figure \ref{Fig1}.i), the number of $\overrightarrow{q}$ vectors
corresponding to the scattering events at the susceptibility
singularity can be symbolized by the area of the intersection
between two CECs translated by $q=2k$. For a circular contour this
area is the same in all directions of the momentum space (as
schematized in e) ), while for the hexagonal one this area is
bigger in the $\overline{\Gamma M}$ direction (blue arrow) than in
the $\overline{\Gamma K}$ direction (red arrow).

\subsection{The joint density of states approximation}

In a previous paper \cite{simonJCondMat07} we have formally
established that the FT of a standing waves pattern image could be
approximated by a joint density of states calculation (JDOS). In
the presence of defects, the quasiparticles in the Bloch state
$\overrightarrow{k}$ can be scattered into the Bloch state
$\overrightarrow{k'}$, and new eigenstates can be constructed, in
perturbation theory, as linear combinations of the degenerate
unperturbed states $\overrightarrow{k}$ that belong to the
constant energy contour $E(k)=\omega$. The leading term in the
Fourier component of
 the LDOS at wave vectors: $\vec q = \vec k' - \vec k
+ \vec G$, where $\vec G$ is a reciprocal lattice vector, has an amplitude which takes the general form
\begin{equation}\label{3}
g(\omega,\vec q) = \frac{1} {{4\pi ^2
}}\!\!\!\!\!\!\iint\limits_{E(k)=E(k')=\omega}\!\!\!\!\!\!\!\!\!\!{f(\vec
k,\vec k',\vec G)\delta (\vec q - \vec k + \vec k' \pm \vec
G})d^{2} \vec k\,d^{2} \vec k'
\end{equation}
where
$f(\overrightarrow{k},\overrightarrow{k'},\overrightarrow{G})$
  is a weighting factor that depends on scattering matrix elements, and thus on the
overall distribution and nature of defects.

The function $g(\omega,\overrightarrow{q})$ gives the joint density of states
with a main contribution
for $\overrightarrow{G}=0$ and replicas shifted by
$\overrightarrow{G}$.
The quantity $g(\omega,\vec q)$ may be calculated by solving the
Schr\"{o}dinger equation for simple-defect geometries. Yet, if many
defects of various symmetries are present, one may assume that
$f(\overrightarrow{k},\overrightarrow{k'},\overrightarrow{G})$ is
a fairly smooth, slowly varying function of $\overrightarrow{k}$
and $\overrightarrow{k'}$.  Thus, practically in the JDOS approximation, $g(\omega,\vec q)$,
which is related to the power spectrum of the LDOS, could be calculated by performing the self correlation function of the CEC at a given
energy. In practice, the JDOS calculations consist in counting the
number of pairs of $\overrightarrow{k},\vec{k}'$ wave vectors, yielding a scattering wavevector $\vec{q}$ with the same length and direction.
We fabricate an image where the
pixels position is defined by the wavevector $\vec{q}$, and the grey contrast
is proportional to the number of the corresponding wavevectors pairs $\vec{k},\vec{k}'$. This is a simple phenomenological
approach of the generalization of the Lindhard
susceptibility to Bloch waves states as developed by Blandin \cite{Blandin}.

\begin{figure}
\begin{center}
\epsfig{file=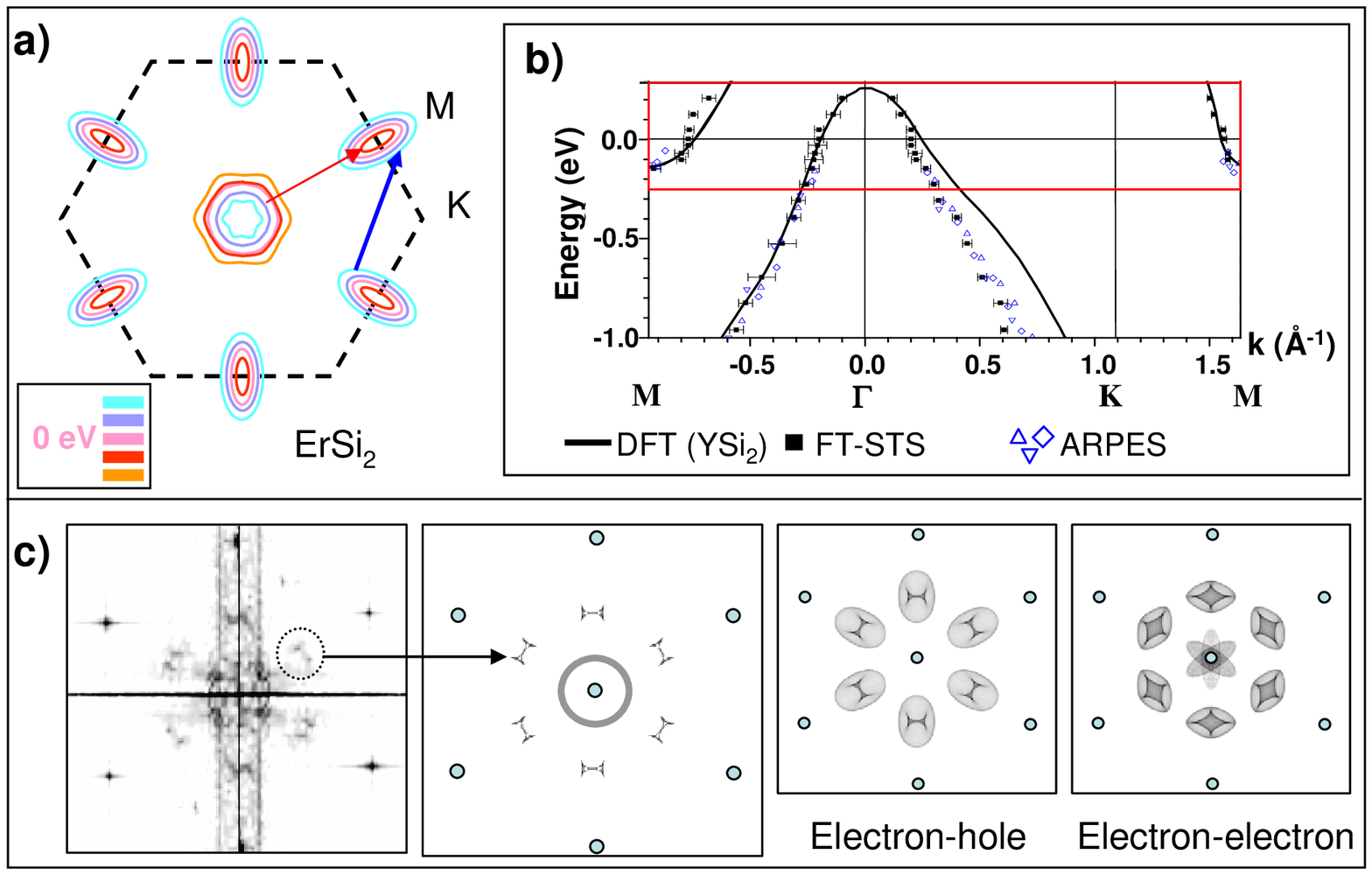,width=8.5cm,angle=0}
\end{center}
\caption{Constant Energy Contours of $ErSi_2$ in the energy region
where both hole and electron band coexist (-200 meV to + 200 meV).
b) shows the experimental band structure of $ErSi_2$ determined
with ARPES and FT-STS measurements. The band structure is compared
to DFT calculation of $YSi_2$. c) from left to right : Fourier
transform of an STM image of $ErSi_2$. The dominating structures
observed in M can be modelized by JDOS construction as depicted in
Fig \ref{Fig1}. JDOS calculations showing the contribution of
different scattering events: the third frame shows the scattering
events between the hole-like band and the electron-like band
whereas the fourth frame displays the structure obtained by only
electron-like band scattering. }\label{Fig2}
\end{figure}

The figure \ref{Fig2} illustrates the fact that not only the
structure, but also the shape of the features observed in the power
spectrum, could be interpreted with a simple JDOS approach. In a)
and b) are represented the CECs, the measured and
theoretical band dispersion of the $ErSi_{2}$ system respectively.
Up to -250meV (rectangular red box in b)), the CEC is split into
two types of contours, the hole-like band structure previously
discussed in Fig.\ref{Fig1} and, crossing the Fermi level from
unoccupied states, six ellipsoidal electron pockets emerging
around the $M$ points. In the JDOS interpretation of the
FTs, we should consider both the scattering  processes  joining hole-like
and electron-like pockets (red arrow in a)), as well as scattering
between the ellipsoidal electron-like pockets (blue arrow). Both
should yield features centered around the $M$ points.

By changing the bias voltage in this energy range, the position of
the feature does not change but only the shape. In c) we show the
FT obtained at the Fermi level. One of the features arising in the
FT is highlighted, and we can see that it has a "butterfly" shape.
We have found that this particular shape is associated to the self
correlation between two CECs, a central circle (the hole band) and
an elliptic electron pocket, i.e. it is the result of the
scattering events schematized by the red arrow in a). The JDOS
calculation for electron-electron scattering is also showed in c).
This clearly demonstrates that the leading scattering event joins
hole-like and electron-like pockets.

The dispersion relation as obtained from the FT-STS measurements
is reported in b). This relation was deduced from the modification
of the feature size with the bias voltage \cite{VonauPRL05}. An
excellent agreement between our experimental points and previous
ARPES measurements \cite{WetzelSSCom92} is found. The agreement is
also excellent with the calculated band structure from Rogero et
al. \cite{RogeroPRB04} in the direction $\overline{\Gamma M}$, but
not in the direction $\overline{\Gamma K}$. Obviously, the
$YSi_{2}$ DFT-LDA calculation reproduces the global shape of the
measured hole band but the higher-energy excitations are shifted
by values as large as $250$ meV. Similar shifts for the predicted
surface-state energy positions around the $K$ point have been
mentioned before for $YSi_{2}$ and $GdSi_{2}$ and the necessity to
improve LDA self-energy term has been noted \cite{RogeroPRB04}.

The case of similar dispersing CECs for high $T_{C}$
superconductors has been studied in detail (see e.g.
\cite{McElroyNature03, Vershinin2004}), and more recently, for
$Bi_{2}Sr_{2}CaCu_{2}O_{8+\delta}(Bi2212)$, the relationship
between the ARPES measurements and the quasiparticule scattering
interpretation of FT-STS measurement has been done using the
concept of JDOS \cite{MarkiewiczPRB2004, K. McElroyPRL2006}. A
more thorough theoretical approach to interpret these results,
based on a Green function formalism, has been proposed by Wang and
Lee \cite{QiangPRBR03}. Thus one can link the two measurement
techniques yielding the \textit{real space single particle
spectra} for the STM and LDOS measurements,
 and the \textit{momentum space single particle spectral function} for the ARPES measurements \cite{K. McElroyPRL2006}.
 To summarize, as the first order stationary
perturbation theory is invoked in the JDOS approach, and as large
signal-to-noise ratio is expected with this technique it is
crucial to apply it to physical systems in which the scattering
processes mix sufficiently the unperturbed wave functions in
k-space and in energy. It has been discussed in detail by
Capriotti et al. \cite {capriottiPRB03}, by Kodra and Atkinson
\cite{KodraPRB06}, and for the case of high $T_{C}$
superconducting materials \cite{HoffmanScience02,McElroyNature03}.
We note that this technique has been successfully applied even
more recently for $GaN(0001)$ \cite{Sun2010}.

\subsection{Beyond the JDOS approximation - the stationary-phase approximation}

Besides the electron-hole scattering processes, one should also take into
 account the hole-hole and electron-electron scattering events. At first glance, the predominance
  of electron-hole scattering events seems to be the effect of hidden matrix elements included in the function
$f(\overrightarrow{k},\overrightarrow{k'},\overrightarrow{G})$ in
the expression of the JDOS (equation \ref{3}). However we can show that
a simple interpretation can already be provided using
the background theory of the susceptibility.

\begin{figure}
\begin{center}
\epsfig{file=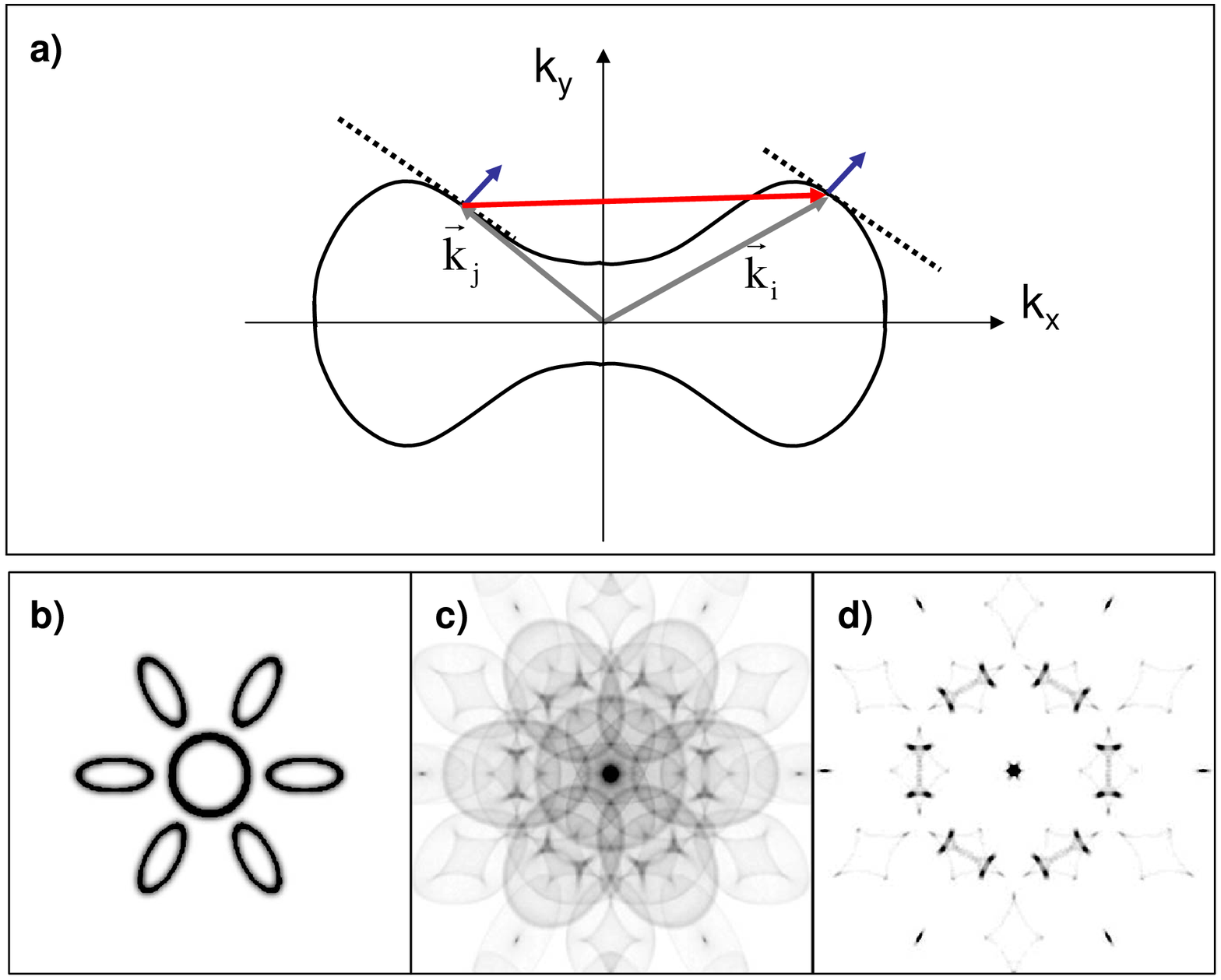,width=8cm,angle=0}
\end{center}
\caption{a) Representation of the stationary phase (blue parallel
vectors) of a scattering event (red vector) resulting from the
scattering from one state to another (depicted by the grey wave
vectors) \cite{Blandin}. b) CEC from erbium disilicide, and in c)
the corresponding JDOS calculation. d) shows the dominating
structures from c), when the stationary phase condition is taken
into account; the result is in good agreement with the
experimental FFT from Fig. \ref{Fig2}.a). }\label{Fig3}
\end{figure}

Up to now we have considered the steady states and neglected the
phase factors in the rough JDOS approximation. In order to take
into account these phase factors, the method of the stationary
phase approximation is usually applied. It consists in noting that
in the integral of the JDOS formula,  oscillatory term with
rapidly-varying phase will cancel, while the terms with the same
phase should be added together. This leads to a supplementary
criterion in the geometrical construction of the JDOS, similar to
the case of the Kohn anomaly. As illustrated in figure
\ref{Fig3}.a) and along the lines of Ref.~\cite{Blandin}, we
consider that the leading scattering process take place for values
of the $\vec{q}$ momenta joining points of the CECs for which
$\overrightarrow{\nabla_{k}(E(k))}$,
$\overrightarrow{\nabla_{k'}(E(k'))}$ (or alternatively the
tangents to the CECs) are parallel and point in the same
direction.

In b) we consider a CEC similar to the Fermi surface of the Erbium
disilicide with the central circle and the six ellipses. The full
calculation of the JDOS is given in c) where we can easily
recognize the features associated to the hole-electron and
electron-electron scattering processes. By applying however the
above-described criterion in our calculation we have obtained the
JDOS given in d). Only the "butterfly" features are preserved in
the resulting spectra. This shows the importance of the phase
factor, as well as that the matrix elements have no real effect on
the FT features. This experimental result also nicely demonstrates
the "stationary phase" mathematical theorem.

\section{FT-STS measurements on graphene and the T-matrix approximation}

\subsection{T-matrix approximation}

The T-matrix approximation has been successfully used to
calculate the effects of disorder on the spectral properties of an
electronic system. The basic theory is described for example in
\cite{Mahan2000, Bruus2005}, and for a few example of how this is applied we
mention Refs.~\cite{Byers1993, Salkola1996, Ziegler1996, Hirschfeld1988, Polkovnikov2001, Podolsky2003, dhl, ddw, benakivelson, benaprl, Wehling2007, Peres2007, Peres2006, Vozmediano2005, Ando2006, Pogolerov, Skrypnyk2006, Skrypnyk2007, Katsnelson2008}.
The basic principle of this technique lies in an infinite
perturbative summation of the diagrams resulting from expanding
perturbatively the Green's function of a system to all orders in
the impurity scattering. In order to be able to apply the T-matrix approximation one needs to have the exact form of the tight-binding Hamiltonian of a system. The T-matrix approximation is valid as long as the impurity potential considered is localized, since it is this condition that allows the re-summation of all orders in perturbation theory. For extended impurities, the T-matrix approximation is in general replaced by the Born approximation, for which only the first order term in the impurity potential series is considered; this is equivalent to a perturbative expansion in the impurity potential which is valid only when the impurity scattering is weak. Nevertheless, both the T-matrix approximation and the Born approximation yield the same dependence of the DOS on position, what is different is the dependence of the DOS with energy, which is not the main point of our analysis.

\subsection{T-matrix approximation in momentum space}

We will briefly review here the principle of the T-matrix
approximation \cite{Mahan2000, Bruus2005}. The impurity scattering problem can
be solved both in the real space and in momentum space. We will
first focus on the momentum space calculation. For a given system
one can define a finite temperature (imaginary time) generalized
Green's function,  \ba G(k_1,k_2,\tau)=-{\rm Tr}\,
e^{-\beta(K-\Omega)}\, {\rm T}_{\tau} \,
\psi_{k_1}(\tau)\psi^+_{k_2}(0), \ea where $K=H-\mu N$,
$e^{-\beta\Omega}={ \rm Tr}\, e^{-\beta K}$, and $ {\rm T}_{\tau}$
is the imaginary time ordering operator. For a
translationally-invariant, disorder-free system the generalized
Green's function defined above is non-zero only if $k_1=k_2$
(momentum is conserved), and the generalized Green's function
reduces to the standard Green's function which depends only on one
momentum. However, the generalized Green's function acquires a
non-zero component for $k_1 \ne k_2$ if the system is
inhomogeneous, such as in the presence of impurity scattering.
This component can be calculated using the $T$-matrix formulation
\cite{dhl,ddw,benakivelson,benaprl}: \ba G(k_1,k_2,i\omega_n)
=G_0(k_1,i\omega_n)T(k_1,k_2,i\omega_n)G_0(k_2,i\omega_n), \ea
where
\begin{eqnarray}
G_0(k,i \omega_n)=(i\omega_n-{\cal H}_k)^{-1}, \label{g0g}
\end{eqnarray}
is the unperturbed Green's function of the homogenous system,
${\cal H}_k$ is the Hamiltonian, and \ba &&T(k_1,k_2,i\omega_n)=
V(k_1,k_2) \nonumber \\&&
+\sum_{k'}V(k_1,k')G_0(k',i\omega_n)T(k',k_2,i\omega_n). \label{t}
\ea

\begin{figure}[htbp]
\begin{center}
\includegraphics[width=5in]{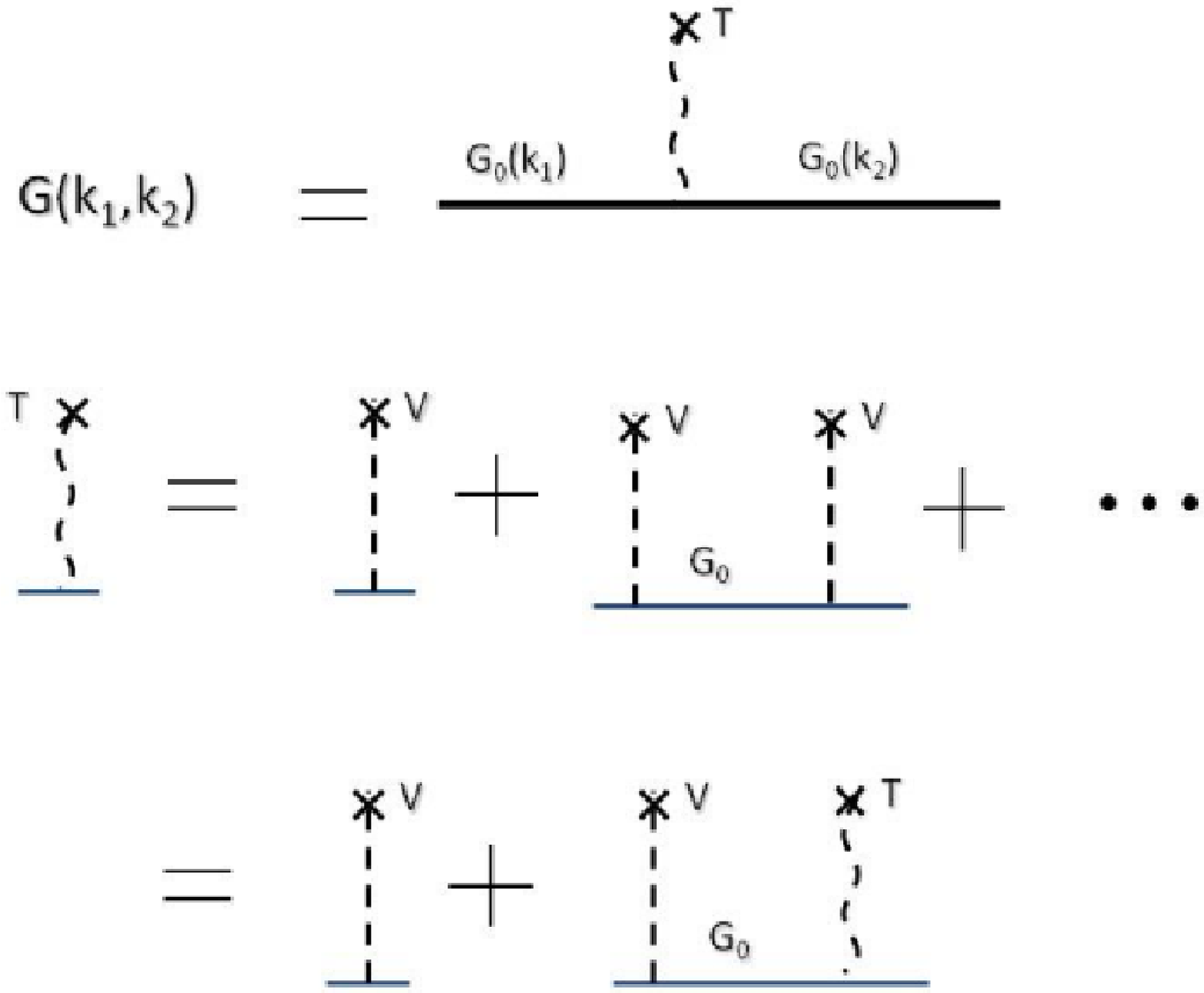}
\vspace{0.in} \caption{\small Diagrammatic description of the T-matrix approximation: the Green's function for a system in the presence of an impurity can be obtained from adding to the unperturbed Green's function the contributions corresponding to all-order impurity scattering processes.
We note that the sum of the contributions appearing to the right of $G_0$ is identical to $T$, which allows one to write the self-consistency equation on the third line.
} \label{fig0}
\end{center}
\end{figure}

As detailed in Fig. \ref{fig0}, this expression stems from a
perturbative expansion to all orders in the impurity scattering.
If only the first term of the expansion is preserved, one recovers
the first order perturbation theory in the impurity scattering
potential, also denoted the Born approximation.
Oftentimes, one assumes that the impurity scattering potential is
very close to a delta function so that $V$ is independent of $k$
and $k'$ \be {\cal H}_{imp}=\int dx V c^\dagger(x) c(x) \delta
(x)=\int_{\vec{k},\vec{k}'} V c^\dagger({\vec{k}}) c({\vec{k}'})
\ee For this case we can solve Eq. (\ref{t}), and obtain
\begin{eqnarray}
T(i\omega_n)&=& [1-V \int \frac{d^2 k}{S_{BZ}} G_0(k,i\omega_n)]^{-1}V
\end{eqnarray}
where $S_{BZ}$ is the area of the first Brillouin zone (BZ) of the
system, and the integral over $k$ is performed over the entire
Brillouin zone.

In the neighborhood of the impurity, spatial oscillations of the
local density of states  are induced. The Fourier transform of
these oscillations can be related to the generalized Green's
function,
\begin{eqnarray}
\rho(q,\omega)=i \sum_{k}g(k,q,\omega), \label{g2}
\end{eqnarray}
where $g(k,q,\omega) \equiv G(k,k+q,\omega)-G^*(k+q,k,\omega)$,
and $G(k,k+q,\omega)$ is the generalized retarded Green's function
obtained by analytical continuation $i \omega_n \rightarrow
\omega+i\delta$ from the Fourier transform of the imaginary-time
Green's function $G(k,k+q, i\omega_n)$.

\subsection{T-matrix approximation in real space}

If one is interested in the real space spectral properties of a
system, such as the space dependence of the LDOS, one can use the
real space T-matrix formalism. The relations described above can
be Fourier transformed to real space, yielding for the retarded
space-dependent generalized Green's function of the system \ba
G(\vec{R}_1,\vec{R}_2,E) =
\int_{\vec{R}}{G}_0(\vec{R}-\vec{R}_1,E) T(\vec{R},E)
{G}_0(\vec{R}_2-\vec{R},E) \label{gr} \ea where $T(\vec{R},E)$ is
the space-dependent $T$-matrix. Same as for the momentum-dependent
formalism, the generalized Green's function depends on two
(spatial) variables. In the absence of disorder, for an
homogeneous system it is only a function of the difference
$\vec{R}_1-\vec{R}_2$, so that only one spatial index is preserved
in the notation: $G_0(\vec{R}_1,\vec{R}_2,E)\equiv
G_0(\vec{R}_1-\vec{R}_2,E)$. However, in the presence of disorder,
the generalized Green's function will depend on both variables
independently.

The spatial dependent Green's function characterizes the
propagation between two points in space $\vec{R}_1$ and
$\vec{R}_2$. If one is not interested in propagation of a particle
between two spatial points, but would rather want to determine the
spectral properties, for example the number of allowed states
having a given energy at a given position, one needs to focus only
on the limit $\vec{R}_1=\vec{R}_2=\vec{r}$. Consequently, the LDOS
can be obtained from the retarded Green's function by using the
conventional relation: \be \rho(\vec{r},E)=-{\rm Im}
G(\vec{r},\vec{r},E) \ee

In general one focuses on a delta-function impurity localized at
$\vec{R}=0$, which yields \ba G(\vec{r},\vec{r},E) =
{G}_0(-\vec{r},E) T(E) {G}_0(\vec{r},E) \label{gr} \ea where
$T(\omega)= [1-V \int \frac{d^2 \vec{k}}{S_{BZ}}
{G}_0(\vec{k},\omega)]^{-1}V,$ and the integral over $\vec{k}$ is
performed on the first BZ, whose area is denoted as $S_{BZ}$.

\subsection{Graphene FT-STS fundamentals}
\label{gft} 
\begin{figure}
\begin{center}
\epsfig{file=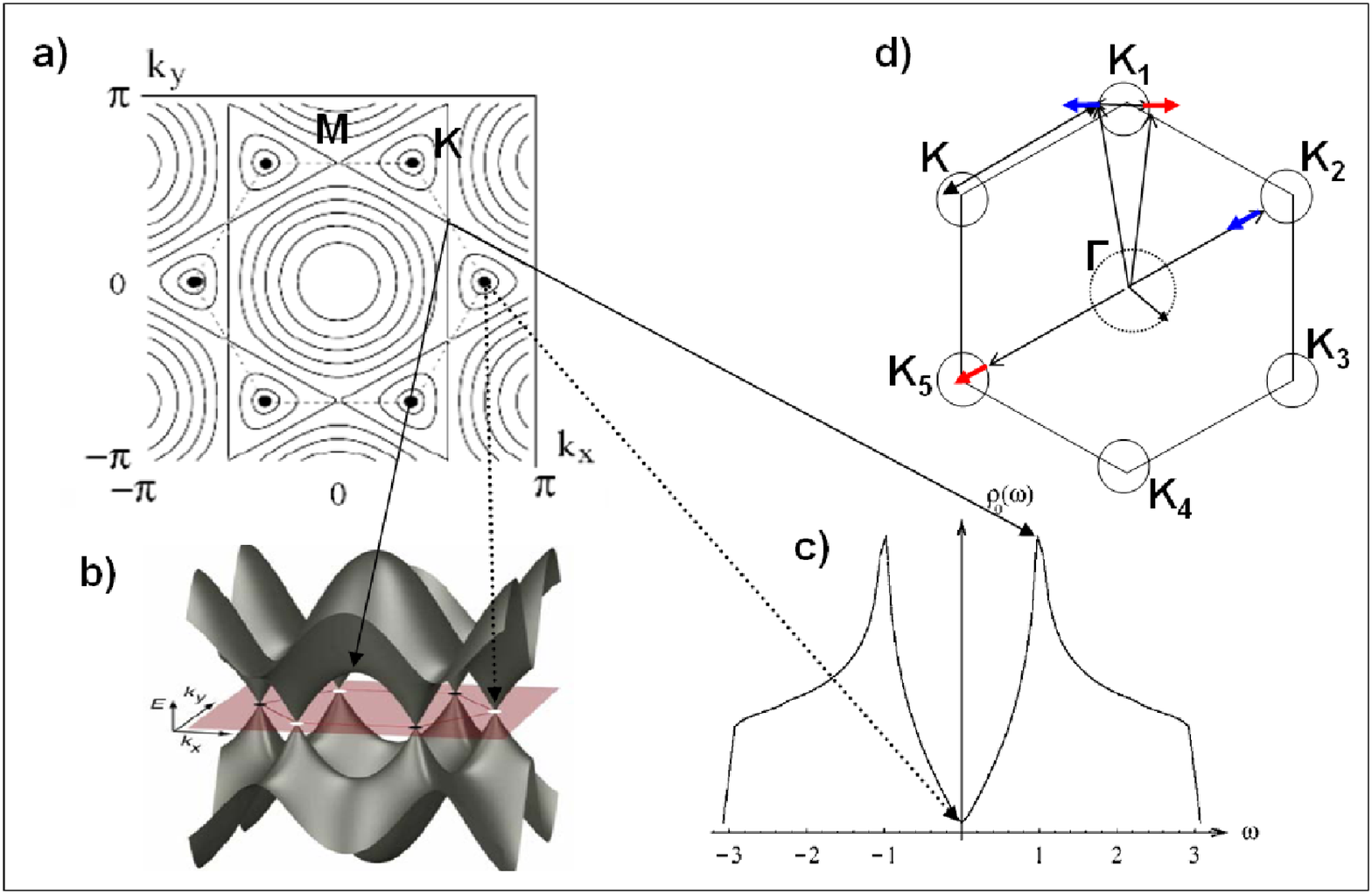,width=9cm,angle=0}
\end{center}
\caption{A three-dimensional (3D) representation of the band structure of graphene, and a plot of the constant-energy contours. a) The constant energy contours starting from the Dirac point, calculated in the nearest-neighbor tight-binding approximation. The key features of the band structure, the Dirac points and the Van Hove singularities are indicated on the 3D representation in b), and on the theoretical density of state in c). The expected intra- and inter-valley scattering processes and the corresponding pseudospin orientations (as discussed in the text) are schematized in d)}
\label{Fig4}
\end{figure}

The Fourier-transform spectroscopy has begun to be
widely used also for graphene systems. Beyond the fascinating
properties of this system and the richness of the possible
fundamental studies, graphene is for us an interesting system
particularly for its band structure and for the CEC contour
topology at specific energies, which provide a good testing ground
for the validity of the FT-STS technique. Figure \ref{Fig4}
recalls the key feature of the band structure of graphene. The
CECs (given in a)) start from Dirac points around K points as
circular contours which become triangular with increasing energy
until the contours touch together at the $M$ points at the so
called Van Hove singularities (VHs). These points are indicated in
the CEC map, on the 3D view of the band structure (given in b)),
and on the corresponding density of states. In d) we have
schematized the CEC map for an energy near the Dirac point and the
possible scattering events expected in the presence of impurities.

Graphene layers have two inequivalent atoms par unit
cells. This is one of the key properties of the graphene. For this
reason graphene should be described by two Bloch waves families,
and the K points are inequivalent (non-equivalent wave vectors
denoted K and K'). Around these K points, and close to zero energy,
it has been showed that an effective Hamiltonian could be defined
as \cite{Castroneto2009}:\[ H_K  = v_F \mathbf{\sigma} .\mathbf{p}
\] where
$\mathbf{\sigma}$, denoted pseudospin, is an operator which
generates the transformation $H_{K}=-H_{K'}$. The pseudospin
provides a supplementary quantum number defined only at low energy
in the immediate vicinity of the K points. The pseudospin is
symbolized by the blue and red arrows in d). As previously
discussed for Erbium disilicide, the expected allowed scattering
$\vec{q}$ wave-vectors could join two points of the same
isocontour (intra-valley scattering) leading to a circle centered
at $\Gamma$, or join two different valleys (inter-valley
scattering). The first type of scattering leads to long wavelength
oscillations in real space, while the second one leads to the well
known short wavelength reconstruction observed in the vicinity of
point defects in graphene, and to the $\sqrt{3}\times\sqrt{3}$
reconstruction oftentimes reported for graphite and graphene
around point defects \cite{RuffieuxPRL00,RuffieuxPRB05} and more
recently near step edges \cite{DujardinNanolett10}.

However, as we will describe in more details in the next section,
not all scattering processes are allowed for graphene. Since the
pseudospin acts as an extra quantum number, two waves of opposite
pseudospin cannot interfere and generate standing waves. Thus,
since for two points on a CEC having opposite momenta relative to
a K point, the pseudospins are opposite, the intra-valley
backscattering is expected to be suppressed. However this is not
the case in a backscattering event $\overrightarrow{K}$,
$\overrightarrow{-K}$ connecting two different opposite K valleys.
In the following we will see how the FT-STS technique could be
useful to probe and analyze the quantum nature of the
quasiparticles on graphene, and how the interpretation of the FT
can be done using the T-matrix approximation.

\subsection{The Hamiltonian and Green's functions for monolayer and bilayer graphene}
The momentum-space tight-binding Hamiltonian for monolayer
graphene \cite{graphene} is: \be {\cal H}=\int d^2 \vec{k}
[a_{\vec{k}}^{\dagger} b_{\vec{k}} f(\vec{k})+h.c.], \label{h0}
\ee where the operators $a^{\dagger}$, $b^{\dagger}$ correspond to
creating electrons on the sublattice $A$ and $B$ respectively, and
$f(\vec{k})=-t \sum_{j=1}^3 \exp(i \vec{k} \cdot {\vec{a}_j})$.
Here $\vec{a}_1=a(\sqrt{3} \hat{x}+\hat{y})/2$,
$\vec{a}_2=a(-\sqrt{3} \hat{x}+\hat{y})/2$, $\vec{a}_3=-a
\hat{y}$, $t$ is the nearest-neighbor hopping amplitude, and $a$
is the spacing between two adjacent carbon atoms, which we are
setting to $1$.

This form of the Hamiltonian has been used in Ref.~\cite{benaprl}
to perform a numerical analysis of the FT-STS spectra. It is also
useful to expand the Hamiltonian close to the corners of the BZ,
which we also denote as nodes or ``Dirac points'', and use the
linearized form to solve the problem analytically at low energies
\cite{benaprl}. The momenta of the six corners of the Brillouin
zone are given by $\vec{K}_{1,2}=[\pm 4 \pi/(3 \sqrt{3}),0]$,
$\vec{K}_{3,4}=[\pm 2 \pi/(3 \sqrt{3}),2 \pi/3]$,
$\vec{K}_{5,6}=[\pm 2 \pi/(3 \sqrt{3}),-2 \pi/3]$. Close to each
corner, $m$, of the BZ we can write $f(\vec{q}+\vec{K}_{m})\approx
\tilde{\phi}_{m}(\vec{q})=v_m \vec{q}\cdot\vec{J}_m$, where
$\vec{q}$ denotes the distance from the respective corner. Also
$v_{1,2}=3 t/2=v$, $v_{3,4}=v \exp(-i \pi/3)$, $v_{5,6}=v \exp(i
\pi/3)$ and $\vec{J}_{1,2}=(\pm 1, -i)$,
$\vec{J}_{3,4}=\vec{J}_{5,6}=(\pm 1,i)$.

The corresponding Green's function, ${\cal G}(\vec{k},\omega)$,
derived from the tight-binding Hamiltonian in Eq.~(\ref{h0}) can
be expanded at low energy around the six nodes (denoted $m$), and
in the $2 \times 2$ (A,B) sublattice basis can be written as:
\begin{align}
{\cal G}(\vec{k},\omega)\approx G_m(\vec{k},\omega)=\frac{1}{\omega^2-|\tilde{\phi}_m(\vec{k})|^2}%
\begin{pmatrix}
\omega+i \delta & \tilde{\phi}_m({\vec{k}}) \\
\tilde{\phi}_m^*(\vec{k}) & \omega+i\delta%
\end{pmatrix}
\label{g0}
\end{align}
where $\delta$ is the quasiparticle inverse lifetime. The Fourier
transform of the linearized Green's function is given by
\cite{benaprl}:
\begin{align}
G_m(\vec{r},\omega)\propto \omega%
\begin{pmatrix}
H_0^{(1)}(z) & i \phi_m(\vec{r}) H_1^{(1)}(z)  \\
i \phi^{*}_m(\vec{r}) H_1^{(1)}(z) & H_0^{(1)}(z)%
\end{pmatrix}
\label{g1}
\end{align}
where $z\equiv \omega r/v $, $H_{0,1}^{(1)}(r)$ are Hankel
functions, $r=|\vec{r}|$, and $\phi_m(\vec{r})=v_m \vec{r}\cdot
\vec{J}_m/(v r)$.

On the other hand, the bilayer graphene consists of two graphene
layers stacked on top of each other such that the atoms in the
sublattice $A$ of the first layer occur naturally directly on top
of the atoms in the sublattice $\tilde{B}$ of the second layer
\cite{vanMieghem1992, McCann2006}, with a tunneling coupling of $t_p$.

 In the sublattice basis $(A, \tilde{B})$ this yields
\cite{McCannPRL2006, Nilsson2006}:
\begin{align}
{\cal H}^{bilayer}_m(\vec{k})=%
\begin{pmatrix}
0 & [\tilde{\phi}_m(\vec{k})]^2 \\
[\tilde{\phi}_m^*(\vec{k})]^2 & 0%
\end{pmatrix}
\end{align}
where for simplicity we have set the effective mass of the
quadratic spectrum to $1$. The corresponding Green's function in
real space is given by:
\begin{align}
G_m(\vec{r},\omega)\propto %
\begin{pmatrix}
H_0^{(1)}(z) &  -[\phi_m(\vec{r})]^2 H_2^{(1)}(z)  \\
-[\phi^*_m(\vec{r})]^2 H_2^{(1)}(z) & H_0^{(1)}(z)%
\end{pmatrix}
\label{g1b}
\end{align}
where we denoted $z=r \sqrt{|\omega|}/v$.

\subsection{Calculations of the FT-STS for graphene using the T-matrix}
\label{st}
Ref.~\cite{benaprl} focuses on monolayer graphene, with a
delta-function impurity localized on an atom belonging to
sublattice $A$. In the $(A,B)$ basis the impurity potential matrix
$V$ has only one non-zero component $V_{11}=u$. The T-matrix
formalism presented above can be generalized to graphene for which
the Green's functions and the T-matrix for graphene become $2
\times 2$ matrices, such that \ba G(k_1,k_2,i\omega_n) ={\cal
G}_0(k_1,i\omega_n)T(k_1,k_2,i\omega_n){\cal G}_0(k_2,i\omega_n),
\ea and where
\begin{align}
{\cal G}_0(k,\omega)^{-1}\propto %
\begin{pmatrix}
\omega  +i \delta &  f(k)  \\
f^*(k) & \omega +i \delta%
\end{pmatrix}
\end{align}
and $ T(\omega)= [I-V \int \frac{d^2 \vec{k}}{S_{BZ}} {\cal
G}(\vec{k},\omega)]^{-1}V, $ where $I$ is the $2 \times 2$
identity matrix, and the integral over $\vec{k}$ is performed on
the BZ, whose area is $S_{BZ}=8 \pi^2/3\sqrt{3}$.

\begin{figure}[htbp]
\begin{center}
\includegraphics[width=4in]{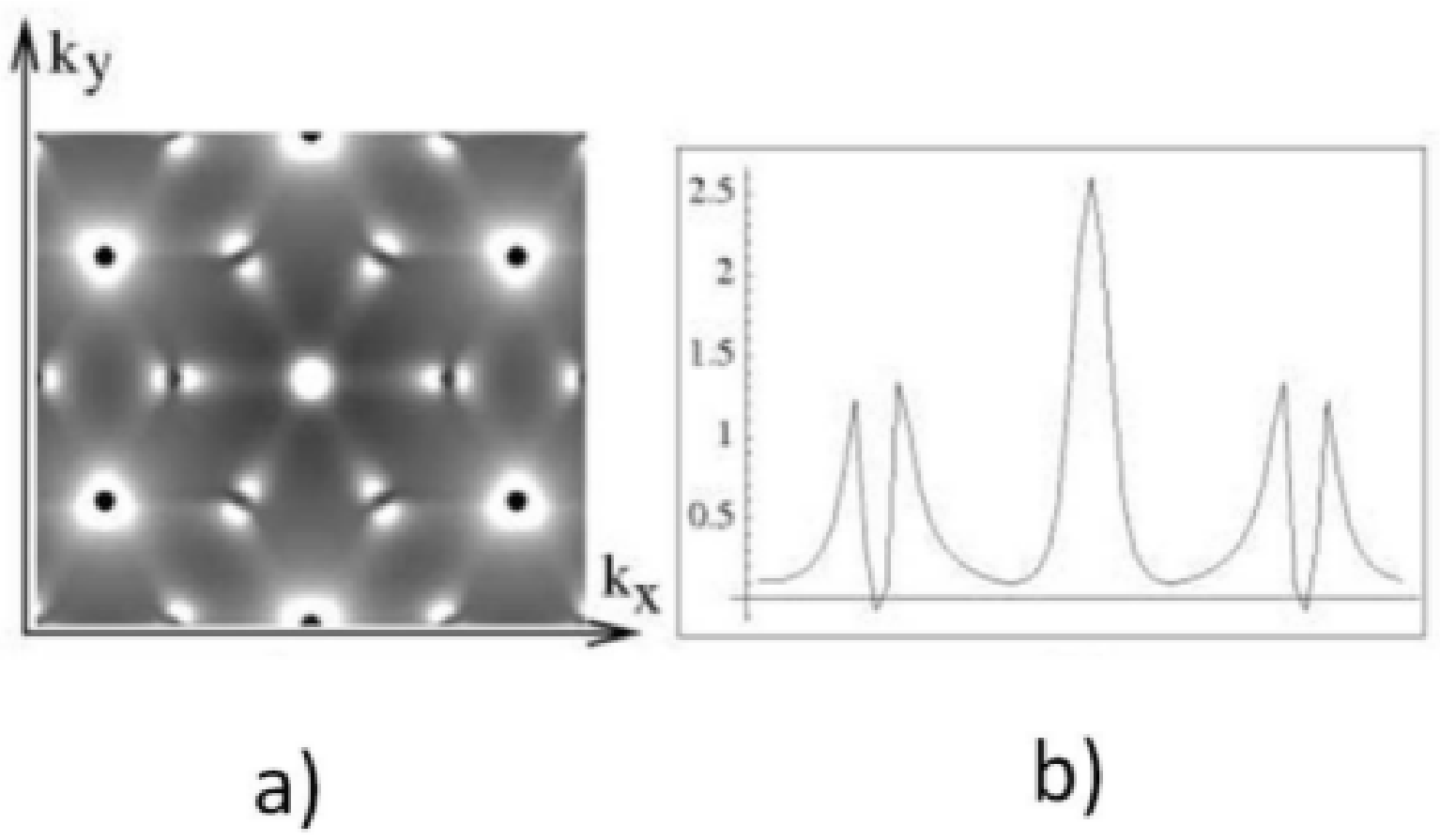}
\vspace{0.in} \caption{\small FT-STS spectra for a monolayer
graphene sample with a single delta-function impurity. Fig. a)
corresponds to energies $0.15t$ at $\delta=0.07t$.  The actual
lowest ($0$) and highest ($1$) values of the FT-STS intensity are
$(-1.3,2.6)$ in arbitrary units. Fig. b) shows a cross section of
the FT-STS intensity as a function of $k_x$ for $k_y=0$ for energy
$0.15t$. } \label{fig1}
\end{center}
\end{figure}

At arbitrary energy this cannot be calculated analytically, but
can be analyzed numerically \cite{benaprl}, and the resulting
FT-STS spectra (corresponding to the real part of the Fourier
transform of the LDOS) are plotted in Fig.~\ref{fig1}.

The calculation shows regions of high intensity in the FT-STS
spectra corresponding to intranodal quasiparticle scattering
(central region) and internodal scattering (outer regions). The
central high-intensity region is a filled circle, while the outer
regions are empty. Also, the rotational symmetry of the
high-intensity regions located at the corners of the BZ is broken.

We can compare this result to the one obtained by the joint
density of states (JDOS) formalism. The JDOS formalism focuses
mainly on the position in k-space of the quasiparticle peaks at a
given energy, and considers that scattering takes place equally
between all quasiparticles living at a given energy. Formally this
is equivalent to writing

\ba \rho(q,i\omega_n) =\int_{k} {\rm Im}\{ {\rm
Tr}[G_0(k,i\omega_n)] \} V {\rm Im} \{ {\rm Tr}[G_0(k +
q,i\omega_n)] \}, \ea

We can see that in this formalism neither the chiral structure of
the Hamiltonian of graphene, nor the phase of the matrix elements
are taken into account, and one focuses solely on the eigenvalues
of the Hamiltonian. The loss of information is evident when in
Fig.~\ref{fig1b} we plot the JDOS result for graphene:

\begin{figure}[htbp]
\begin{center}
\includegraphics[width=15cm]{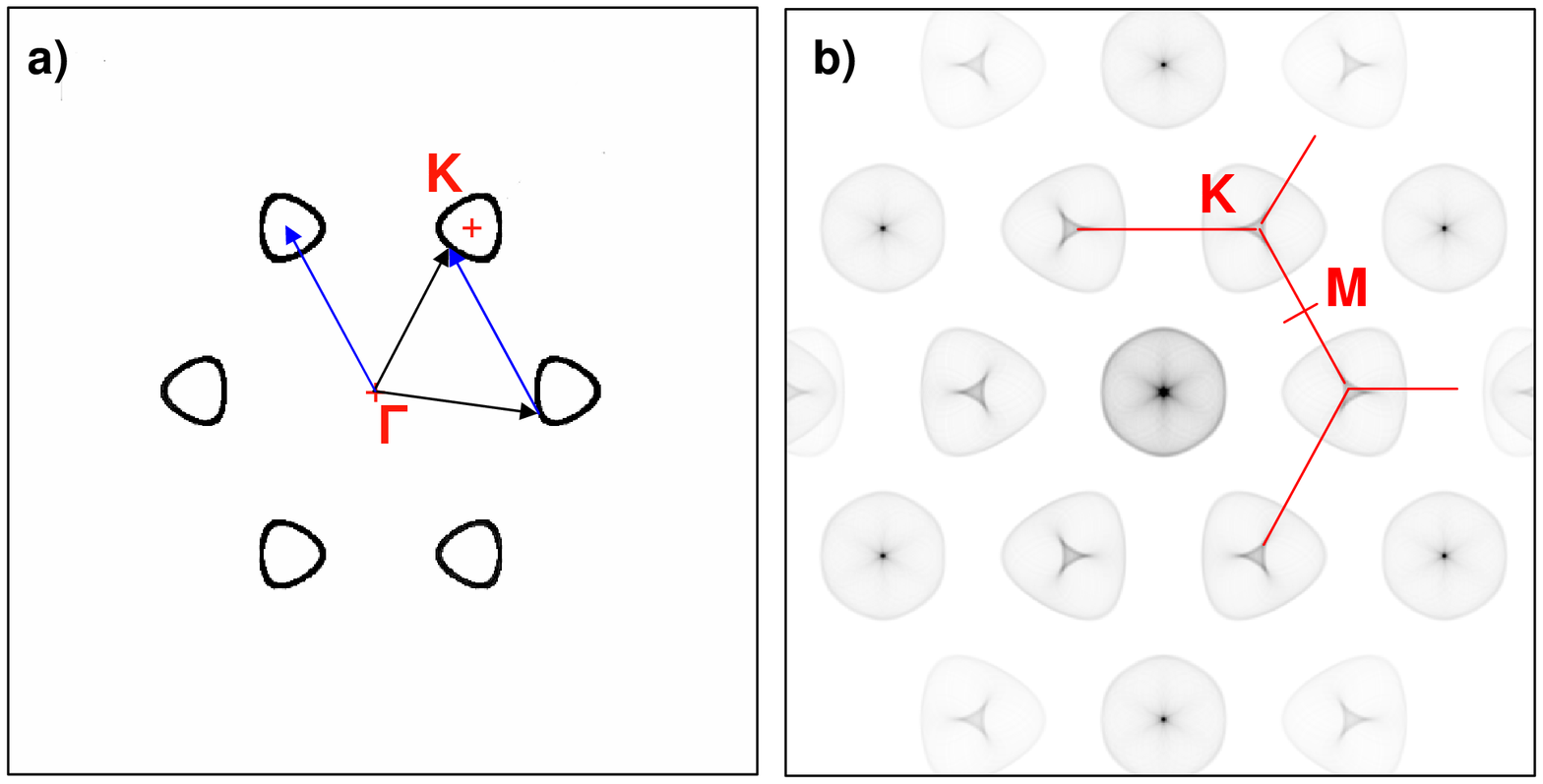}
\vspace{0.in} \caption{\small In b) the FT-STS map for a monolayer graphene calculated using the JDOS approximation for the CEC given in a). The arrows in b) show the inter-valley scattering momenta leading to the feature observed around the K points in b). The central circle corresponding to intra-valley scattering does not appear in the T-matrix calculation, and is not observed experimentally. As detailed in the text, this is a consequence of the chirality of graphene quasiparticles. } \label{fig1b}
\end{center}
\end{figure}

Indeed here the calculated FT-STS shows a central circle
corresponding to the intravalley scattering which is not obtained in the T-matrix calculation,
 nor observed experimentally. The feature observed in the middle of each K
contour (inter-valley scattering) is due to the trigonal warping
of the contour.

\begin{figure}[htbp]
\begin{center}
\includegraphics[width=8cm]{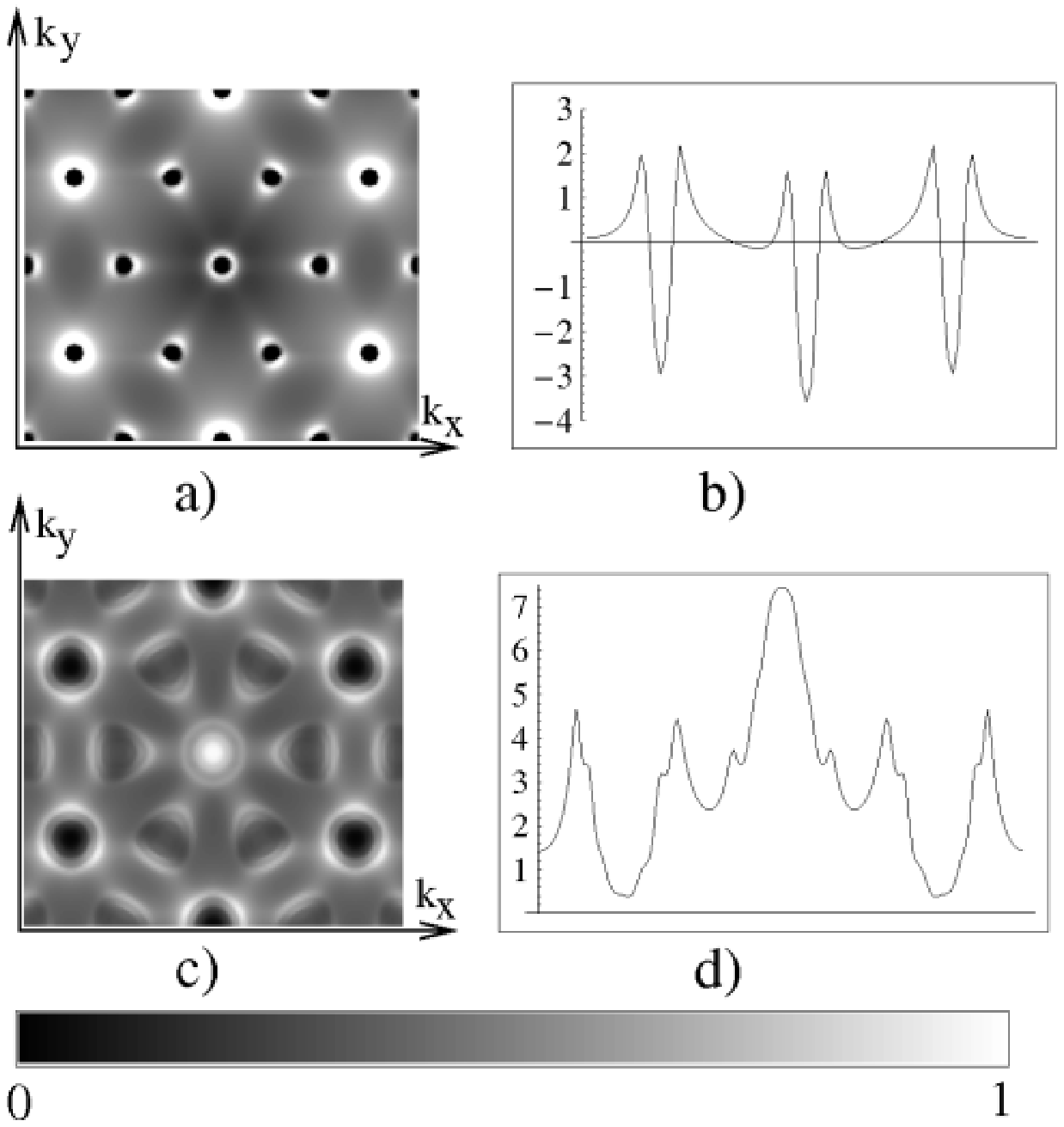}
\vspace{0.in} \caption{\small FT-STS maps for a bilayer sample.
Figs. a) and c) depict the FT-STS intensity in arbitrary units at
energies $0.1t$, $0.4t$, with $t_p=0.3t$, and $\delta=0.05t$. The
actual lowest ($0$) and highest ($1$) values of the FT-STS
intensity are $(-6.9,3.6)$ and $(-6.0,7.4)$ respectively. Figures
b) and d) are cross-sections at $k_y=0$ of Figs. a) and c). }
\label{fig2}
\end{center}
\end{figure}

For bilayer graphene, in Ref.~\cite{benaprl}, one has considered
the case of an impurity located on the sublattice $A$. The
resulting FT-STS spectra for the LDOS in the top layer are
presented in Fig.~\ref{fig2}.

There are similarities and discrepancies between the monolayer and
bilayer cases. Like in the monolayer case, there are areas of high
intensity centered on the center and corners of the BZ. The main
difference at low energy is that the central region of high
intensity is an empty circle, and not a full circle (as for the
monolayer case).  At high energy, we also note a doubling of the
number of high intensity lines corresponding to the doubling of
the number of bands.

\subsection{Calculations of the spatial dependence of the LDOS}

We now turn to the study of the dependence of the LDOS on the
relative position with respect to the impurity ($\vec{r}$). We
note that a calculation of the LDOS at arbitrary energy can be
performed numerically using Eqs.~(12,13), where the real-space
Green's function for graphene can be calculated numerically by
taking a FT of the full k-space Green's function of graphene in
Eq.~(\ref{g0g}). However, at low energies, the physics is dominated
by linearly dispersing quasiparticles close to the Dirac points,
and the calculation can be performed analytically. The spatial
variations of the LDOS due to the impurity have been found to be
\cite{benaprl}: \ba \rho(\vec{r},E) \propto -{\rm{Im}}[{\cal
G}(-\vec{r},E) T(E) {\cal G}(\vec{r},E)]
\approx -\sum_{m,n} {\rm{Im}}[e^{i (\vec{K}_m-\vec{K}_n)\cdot
\vec{r}}G_m(-\vec{r},E) T(E) G_n(\vec{r},E)], \label{gr} \ea where
$m,n$ denote the corresponding Dirac points. Here $T(E)$ is again
the $T$-matrix, which for a delta-function impurity is given by $
T(\omega)= [I-V \int \frac{d^2 \vec{k}}{S_{BZ}} {\cal
G}(\vec{k},\omega)]^{-1}V, $ where $I$ is the $2 \times 2$
identity matrix, and the integral over $\vec{k}$ is performed on
the BZ, whose area is $S_{BZ}=8 \pi^2/3\sqrt{3}$.

Using  Eq.~(\ref{g1}) and expanding the Hankel functions to
leading order in $1/r$, in Ref.~\cite{benaprl}, it has been found
that far from the impurity ($\omega r/v \gg 1$) the corrections to
the local density of states due to scattering between the nodes
$m$ and $n$ are given by: \be \rho_{mn}(\vec{r},\omega) \propto
\frac{\omega}{r}
{\rm{Im}}\big\{t(\omega)e^{i(\vec{K}_m-\vec{K}_n)\cdot \vec{r}+2 i
\omega r/v} i\big[1- \phi_m^{*}(\vec{r})
\phi_n(\vec{r})\big]\big\}. \ee where $t(\omega)$ is the non-zero
element of the $T$-matrix ($T_{11}$), as it was noted that
$\phi(-\vec{r})=-\phi(\vec{r})$.

In the case of intranodal scattering ($m=n$) the above expression
vanishes and the LDOS is dominated by the next leading correction
$ \rho_{m}(\vec{r},\omega) \propto \sin(2 \omega r/v)/r^2$. This
is different from what usual wisdom would suggest for a
two-dimensional system ($1/r$ decay) \cite{benakivelson,steveold},
and has also been described in Refs. \cite{falko,glazman}.

As briefly outlined above in section \emph{Graphene FT-STS fundamentals},
the underlying physics of this result stems from the chirality of
the graphene quasiparticles \cite{Geim,Katsnelson}. The graphene
quasiparticles, due to the presence of two atomic sublattices,
have an additional degree of freedom deemed pseudospin, such that
one says that a quasiparticle belonging to sublattice A has
pseudospin ``up'', and a quasiparticle belonging to the sublattice
B has pseudospin ``down''. By chirality, the pseudospin vector is
parallel to the quasiparticle momentum, so that if we know the
momentum of a quasiparticle, we automatically know its pseudospin,
thus to which sublattice it belongs and in which proportion. Due
to the chiral properties of the quasiparticles, backscattering of
quasiparticles by extended impurities is forbidden
\cite{Ando,Katsnelson2}, because when a quasiparticle is
backscattered flipping its momentum by $180^{\rm o}$, by
chirality, its pseudospin also flips. Since an extended impurity
cannot flip the pseudospin of an electron (the wavelengths
associated with backscattering are much larger than the atomic
lattice constant), the backscattering cannot take place.
Nevertheless, if the impurity is localized, the backscattering is
not forbidden \cite{Katsnelson3}. However, the incident and
backscattered particle have opposite pseudopspins, and thus their
wavefunctions cannot interfere constructively, same as the
wavefunctions of two electrons with spin up and spin down cannot
give rise to constructive interferences. This lack of interference
yields a reduction of the Friedel oscillations in the vicinity of
the impurity, which no longer decay as  $1/r$ as in  a regular
two-dimensional electron gas, but much faster, as $1/r^2$.

We should note that the two-dimensional FT of  $\sin(2 \omega
r/v)/r^2$ is roughly $\rho_m(q,\omega)\propto \pi \theta(2
\omega-q v)/2 +\arcsin(2\omega/q v) [1-\theta( 2 \omega-q v)]$.
This corresponds to a filled circle of high intensity in the FT-STS
spectrum, which is consistent with the results of our numerical
analysis for the central region of high intensity.

Nevertheless, for the decay of the Friedel oscillations generated
by internodal scattering ($m\ne n$), the chirality considerations
are not longer applicable (quasiparticles close to different nodes
have different chiralities), and the leading order behavior of the
Friedel oscillations is $1/r$. The FT of $\cos(2 \omega r/v)/r$ is
$\theta(q v-2\omega)/\sqrt{q^2 v^2-4 \omega^2}$, which translates
into {\it empty} circles of high intensity in the FT-STS spectra,
consistent with our numerical analysis.

For bilayer graphene, an analytical study can be performed at low
energies starting from the expansion of the Hamiltonian around the
Dirac points $m$. Starting from Eq.~(\ref{gr}),
Ref.~\cite{benaprl} has performed a similar analysis to the case
of monolayer graphene. It was noted that at large distances ($z
\gg 1$), as opposed to the monolayer case, the leading ($1/r$)
contribution for intranodal scattering  is non-vanishing: \be
\rho_m(\vec{r},\omega) \propto \frac{1}{r
\sqrt{|\omega|}}\cos(r\sqrt{|\omega|}/v). \ee This is consistent
with the appearance of an empty circular contour at the center of
the BZ, as opposed to the filled circle for the monolayer case.
The leading contribution to the decay of the oscillations due to
internodal scattering is also $1/r$.

\subsection{Experimental measurements of FT-STS in graphene}

\begin{figure}
\includegraphics[width=15cm]{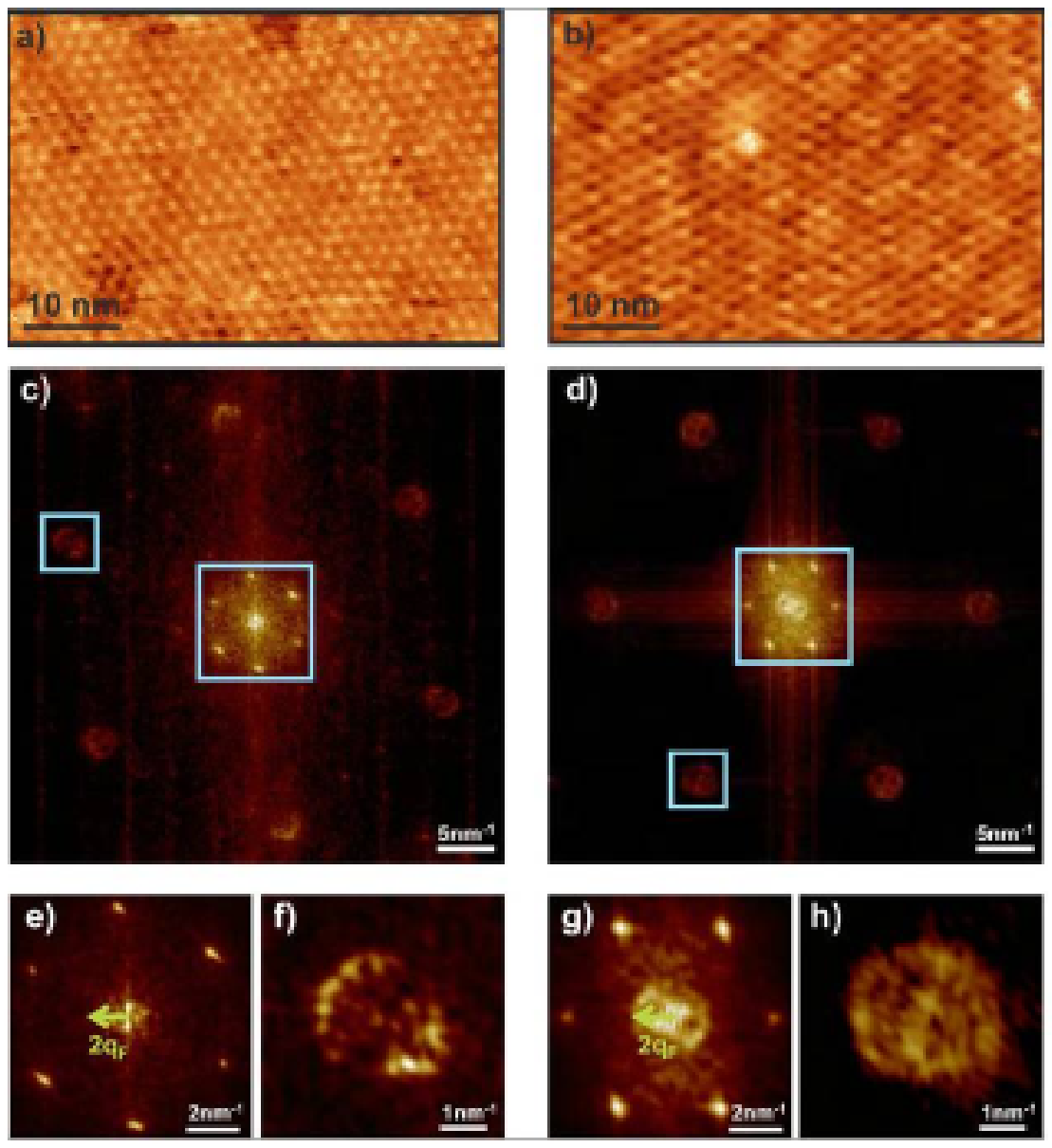}
\caption{a),b) Low-bias STM images of 50 nm wide monolayer a) and
bilayer b) terraces. Sample bias and tunneling current are
respectively +2 mV and 0.4 nA for a), +4 mV and 0.13 nA for b).
c), d) Two-dimensional fast Fourier transform (FFT) maps of the
STM images a) and b). e) Central region of c), showing no
intravalley-backscattering related ring (the green arrow points
out the position where such a ring should appear). f) One of the
outer pockets of c). g) Central region of d), showing a clear
ring-like feature of radius 2$q_{F}$ related to
intravalley-backscattering. h) One of the outer pockets of d).
Outer pockets shown in f) and h) are centered at the $K$ (or $K$')
point and result from intervalley scattering. Data courtesy : I.
Brihuega et al. \cite{benamallet}. } \label{fig3}
\end{figure}

\begin{figure}
\begin{center}
\includegraphics[width=10cm]{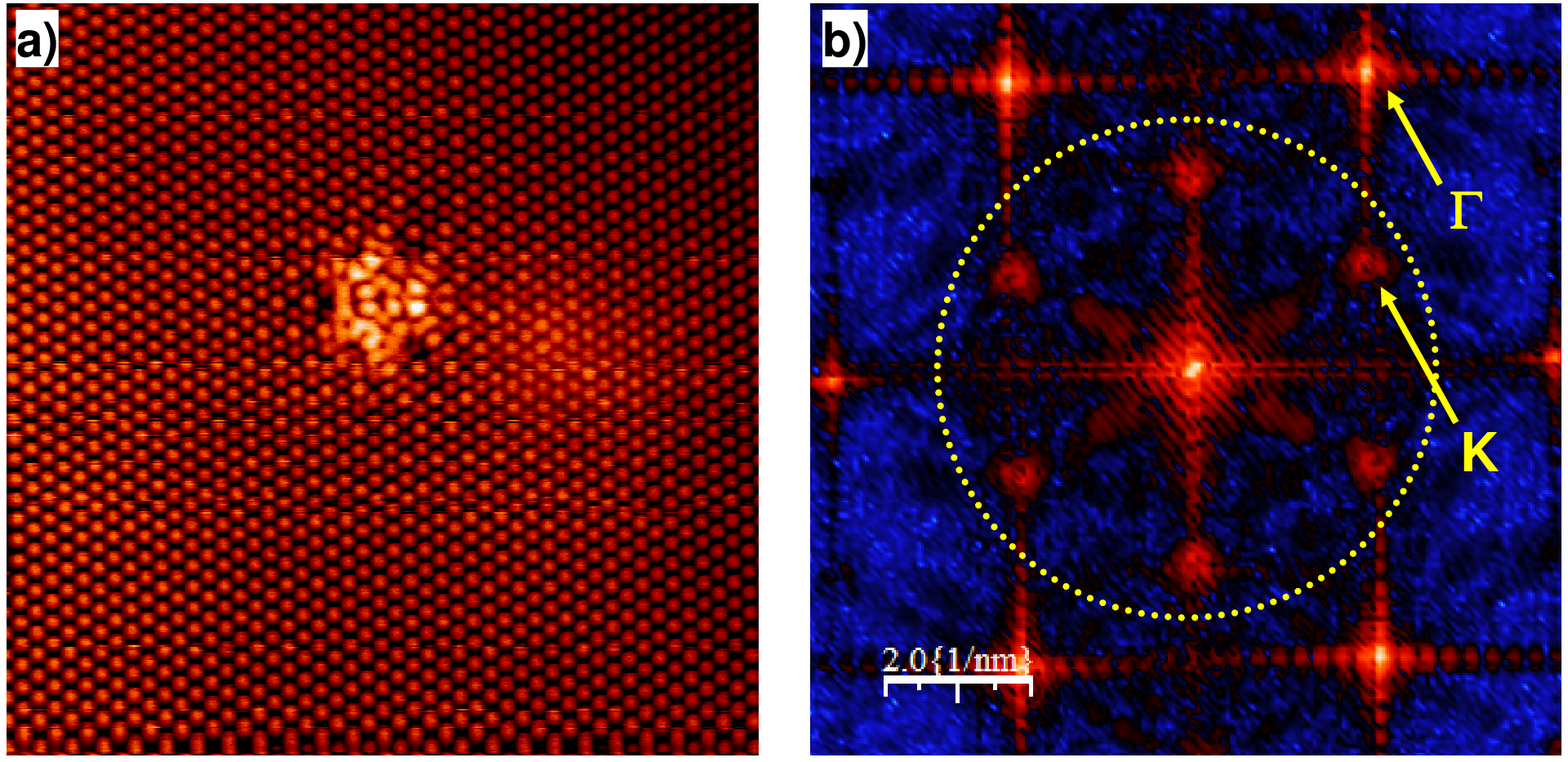}\\
\end{center}
\vspace{0in} \caption{a) STM topographic image ($10\times
10nm^{2}$, -17$meV$, 1$nA$) showing an isolated defect. b) The FFT
power spectrum of the 2D topographic image in a).The features
indicated by the arrows correspond to inter-valley scattering. }
\label{fig4}
\end{figure}

Experiments to measure the LDOS in graphene have been performed by
quite a few groups
\cite{mallet,rutter,benamallet,benasimonEPJB,cranneyEPL10}. Thus,
for example, we present below some real-space images from
Ref.~\cite{benamallet}, for graphene monolayer
(Fig.~\ref{fig3}.a)) and graphene bilayer (Fig.~\ref{fig3}.b)).
Note that both images exhibit a triangular pattern of periodicity
$\sim$ 1.9 nm which is related to the interface reconstruction
\cite{mallet,rutter}, and which appears as a sextuplet of bright
spots in the corresponding FFT images([Figs.~\ref{fig3}.c) and
\ref{fig3}.d)). Also in Fig.~\ref{fig4}, we present the LDOS and
its Fourier transform as measured in Ref. \cite{benasimonEPJB}.

The central region of the FFT in Figs.~\ref{fig3}.c) and
\ref{fig3}.d) is related to intravalley scattering: a clear
ring-like feature of average radius 1.2 nm$^{-1}$ is found for the
bilayer (Figs.~\ref{fig3}.d) and \ref{fig3}.g)). This radius value
is in agreement with the value given in ref. \cite{rutter}, and
with the 2$q_{F}$ value derived from ARPES \cite{Bostwick,Zhou}.
On the monolayer terrace, no central ring is found
(Figs.~\ref{fig3}.c) and \ref{fig3}.e)), despite the unprecedented
momentum resolution (the result has been checked on many different
monolayer terraces). This corresponds to the presence of
slowly-decaying ($1/r$) long-wavelength oscillations in bilayer
graphene, but not in monolayer graphene.

In the high frequency regions of the FFTs of Figs. \ref{fig3}.c)
and \ref{fig3}.d), six outer pockets with ring-like shapes
centered at $K$($K'$) points have been observed. They result from
intervalley scattering, associated to real space LDOS modulations
with a ($\sqrt{3}\times\sqrt{3})R30^{\circ}$ periodicity with
respect to graphene \cite{mallet,rutter}. As shown in
Figs.~\ref{fig3}.f) and \ref{fig3}.h), the intensity of the high
frequency rings in the FFT is not isotropic. The anisotropy is
much more pronounced for graphene monolayer. The presence of the
anisotropy is in agreement with the theoretical calculations of
Ref.~\cite{benaprl} presented in Fig.~\ref{fig1}.

The experiment described in \cite{benamallet} thus confirms the
theoretical picture presented in section \emph{Calculations of the FT-STS for graphene using the T-matrix}, confirming the
chiral properties of the graphene quasiparticles, and also that
these properties can be probed at the nanometer scale using
scanning tunneling spectroscopy.

\subsection{Determination of the band structure of graphene from the FT-STS spectra}
\begin{figure}
\begin{center}
\includegraphics[width=15cm]{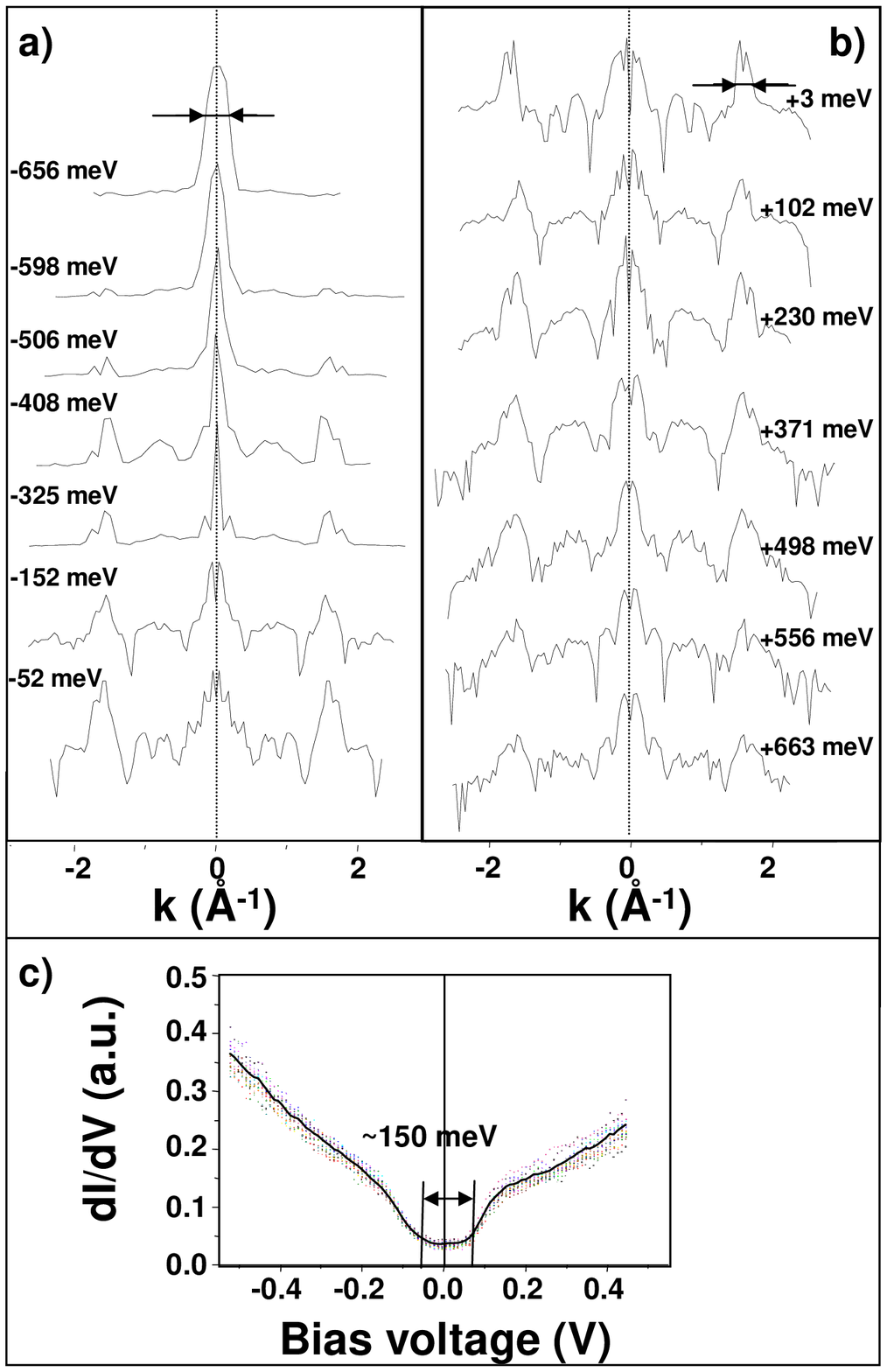}\\
\end{center}
\vspace{0in} \caption{a), b) Intensity profiles along the
$K-\Gamma-K$ direction of the 2D FT of the LDOS taken for
different bias voltages. c) STS measurement taken near the
impurity showing a gap centered around the Fermi level.
}\label{fig5}
\end{figure}

The Fourier transform of the LDOS can be used not only to extract
information about the chirality of quasiparticles, as described
above, but also about the quasiparticle energy dispersion (see for
example \cite{McElroy2005, Vershinin2004, Fang2004, rutter}). In
Fig.~\ref{fig5} we present intensity profiles along the
$K-\Gamma-K$ direction for the Fourier transform of the LDOS at
various energies \cite{benasimonEPJB}. Every image was acquired
using a lock-in amplifier and a modulation voltage of $\pm 20$
$meV$ which gave the energy uncertainty. For a wide range of
energies, between energies well below the Fermi level ($-950$
$meV$) to about $-150$ $meV$, the width of the peak at the center
of the Brillouin zone (measured at half maximum) shows a clear
dispersion with the bias voltage (it decreases when the energy
moves towards the Fermi level). At energies closer to the Fermi
level the profile becomes more complex and displays a dip in the
intensity profile close to the $\Gamma$ point. The central peak
also shows two lateral structures (shoulders). The width of the
central feature does not appear to disperse with energy close to
the Dirac point, but it starts dispersing again for higher
positive energies.

\begin{figure}
\begin{center}
\includegraphics[width=15cm]{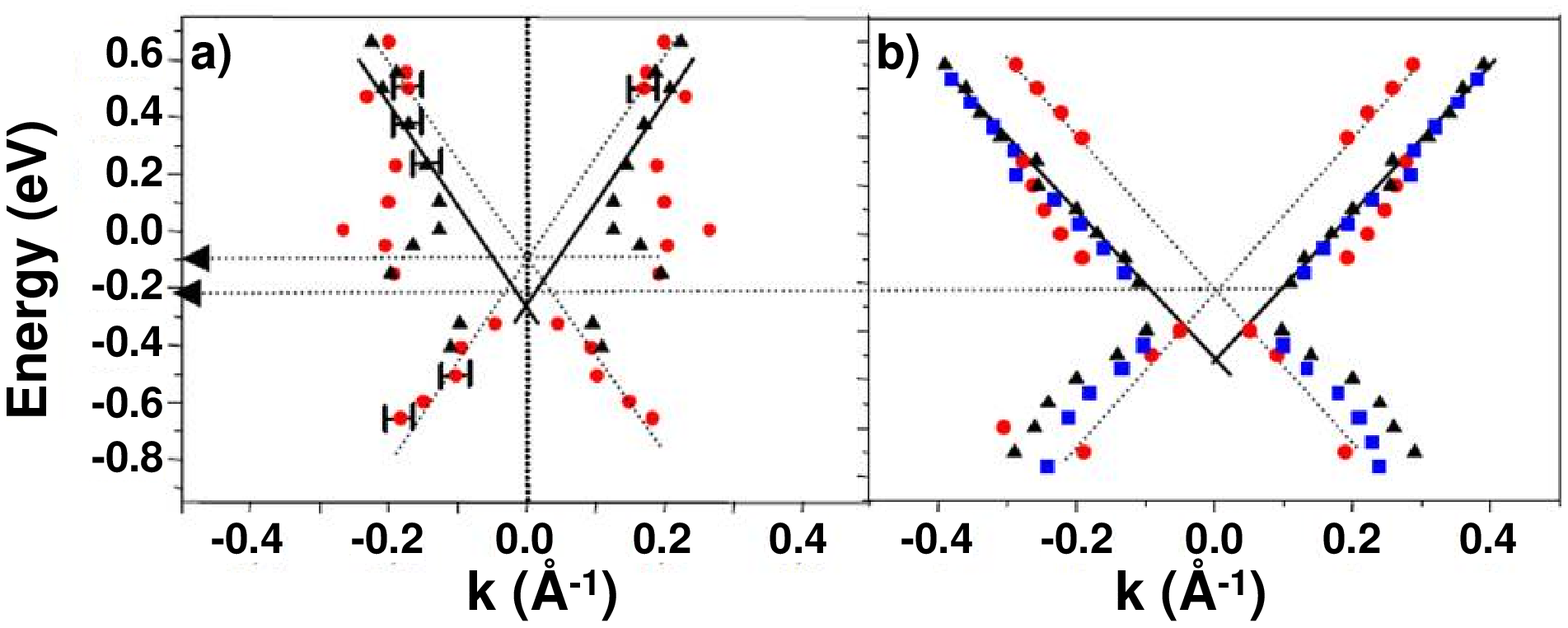}\\
\end{center}
\vspace{0in} \caption{The dispersion for the width of the
central-ring feature (red circles) and of the K-points feature
(black triangles) obtained in a) from the experimental data from
Fig. \ref{fig5}, and from the theoretical $K-\Gamma-K$ cuts (b)).
The dispersion of the central-feature shoulder on the theoretical
curves is indicated by blue squares. }\label{fig6}
\end{figure}

Fig.~\ref{fig6} shows a linear dependence of the width of the
central feature with energy, except in an energy range from $-200$
$meV$ to $100$ $meV$. For these energies the shape of the central
peak is more complex. Also, in this range of energy,
Fig.~\ref{fig5}.c) presenting the measurement of the LDOS as a
function of energy taken close to the impurity, shows a $150$
$meV$ gap-like feature centered around the Fermi level at $0$
$meV$. The Dirac point is estimated at $-100$ $meV$ below the
Fermi level. The spreading in the dispersion of the points close
to the Fermi level could be attributed to the presence of the gap.

The width of the features at the K points follows a similar
dispersion. Beyond $+500 meV$, all the features seem to follow a
different dispersion branch, as indicated in Fig.~\ref{fig6}. In
ref.~\cite{benasimonEPJB} a comparison between the experimental
results with similar theoretical profiles obtained using a
T-matrix approximation for a single localized impurity in bilayer
graphene \cite{benaprl} is also presented. No gap at the Fermi
level was included in the theoretical calculation, while a small
gap of $\approx 100$ $meV$ was assumed near the Dirac point, which
was taken to be close to $-250$ $meV$. In Fig.~\ref{fig6}.b) is
plotted the dispersion of the central and K-points features, and
of the shoulder, as obtained from the theoretical curves. One
observes the presence of two different dispersion branches. The
second dispersion branch and the shoulder arise because of the
bilayer graphene upper band which opens at energies higher than
the inter-layer coupling. For monolayer graphene these extra
features should not appear.

Thus the measurements in \cite{benasimonEPJB} show that the
quasiparticle approximation and the Fermi liquid theory are robust
over a large range of energies. The complex structure of the
central feature (the existence of the shoulder), as well as the
presence of two dispersion branches are consistent with the
bilayer (or multilayer) character of the graphene sample. While
the point defect modifies the electronic wave-function in its
vicinity, a clear linear dispersion is still observed, and the
relativistic character of the quasiparticle is preserved. Also,
the STS measurements in Fig.~\ref{fig5}.c) indicate the presence
of a gap centered at the Fermi point, and not at the Dirac point
inferred from our FT-STM measurements. This gap is thus different
from the Dirac-point gap observed by ARPES \cite{Bostwick,Zhou}
which was attributed to the different doping levels of the
epitaxial graphene layers. It is however consistent with previous
STM measurements performed on epitaxial graphene \cite{BrarAPL07},
and more recently on exfoliated graphene \cite{Crommie08}, where a
gap observed at the Fermi level was attributed to a pinning of the
tunneling spectrum due to the coupling with phonons.

\subsection{Measurement of the spatial dependence of the LDOS in the vicinity of a defect}
\begin{figure}
\begin{center}
\includegraphics[width=15cm]{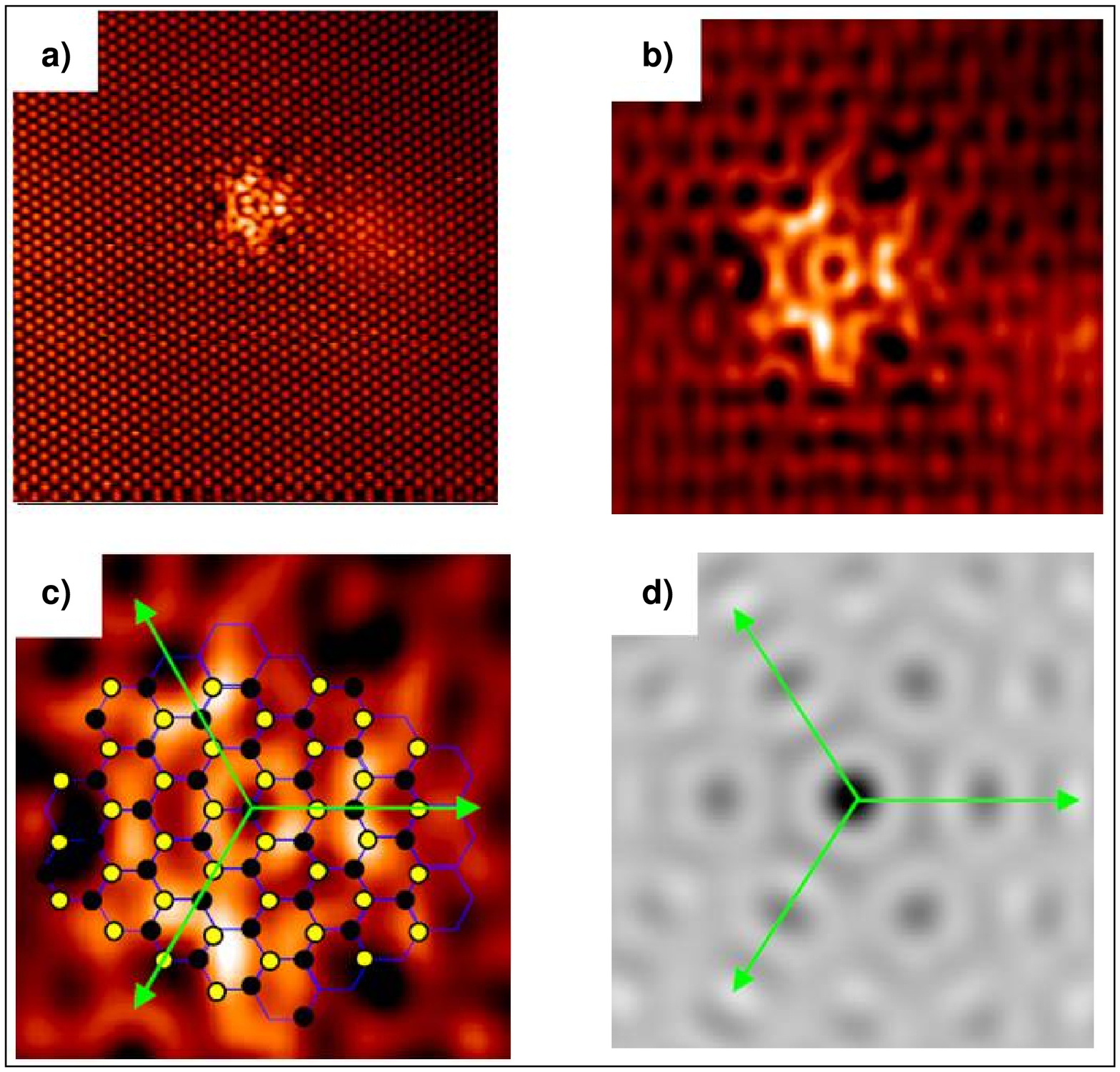}\\
\end{center}
\vspace{0in} \caption{a) STM topographic image ($10\times 10
nm^{2}$, -17$meV$, 1$nA$) showing an isolated defect. b) Zoom-in
($5\times 5 nm^{2}$) on the defect, subsequent to a FFT filtering
removing atomic resolution and second order features in the 2D
FFT. c) $2.5\times 2.5 nm^{2}$ zoom in on b) with the schematic
atomic lattice of graphene (A/black and B/yellow atoms)
superimposed over the standing waves pattern around the point
defect. d) The real-space LDOS in a bilayer graphene calculated
for a point defect using a single-impurity T-matrix approximation
at $300meV$ above the Dirac point. }\label{fig7}
\end{figure}

Figure \ref{fig7}.a) shows a topographic image of a large graphene
terrace taken at $-17$ $meV$ (probing filled states), as measured
in Ref.~\cite{benasimonEPJB}. This layer shows an intriguing
``star-like'' defect with an apparent six-fold ($C6v$) symmetry.
This atomic defect is accompanied by a strong distortion of the
graphene lattice. The center appears black, which is a dramatic
change from the case of the unperturbed lattice.  As schematized
in figure \ref{fig7}.c), a detailed analysis of the real-space
image, shows that the point defect is directly located above or
below a lattice site.

Using FFT filtering, Ref.~\cite{benasimonEPJB} has removed the
lattice periodicity vectors and all other features with
wavevectors outside the yellow circle in Fig.~\ref{fig4}.b), thus
taking into account only the intra-valley and the
inter-consecutive-valley scattering processes. The resulting
real-space image is shown in figure \ref{fig7}.b). This operation
strongly enhances the anisotropic intensity observed also on the
bare topographic image. The LDOS near the defect shows a clear
threefold ($C3v$) symmetry. While, as depicted in
Fig.~\ref{fig7}.c), close to the impurity one observes a fairly
homogeneous standing waves ring, with an almost perfect six-fold
symmetry, farther away from the impurity, the intensity is clearly
higher along three axes (drawn in green in Figs.~\ref{fig7}.c) and
\ref{fig7}.d)).

In Ref.~\cite{benasimonEPJB} the possible origin of this
three-fold symmetry has been analyzed by comparing the theoretical
LDOS for a point defect to the experimental data. The theoretical
LDOS modulation for a bilayer graphene in the presence of a single
impurity has been calculated using the real-space T-matrix
formalism, modified to incorporate the finite spatial extend of
the carbon electronic orbitals \cite{benaprb}. The results for an
energy of $300$ $meV$ above the Fermi level  are presented in
Fig.~\ref{fig7}.d). One notes quite a few features similar to the
ones depicted in Fig.~\ref{fig7}.c), including the existence of a
three-fold symmetry. However the three-fold anisotropy is much
less pronounced.

\begin{figure}
\begin{center}
\includegraphics[width=15cm]{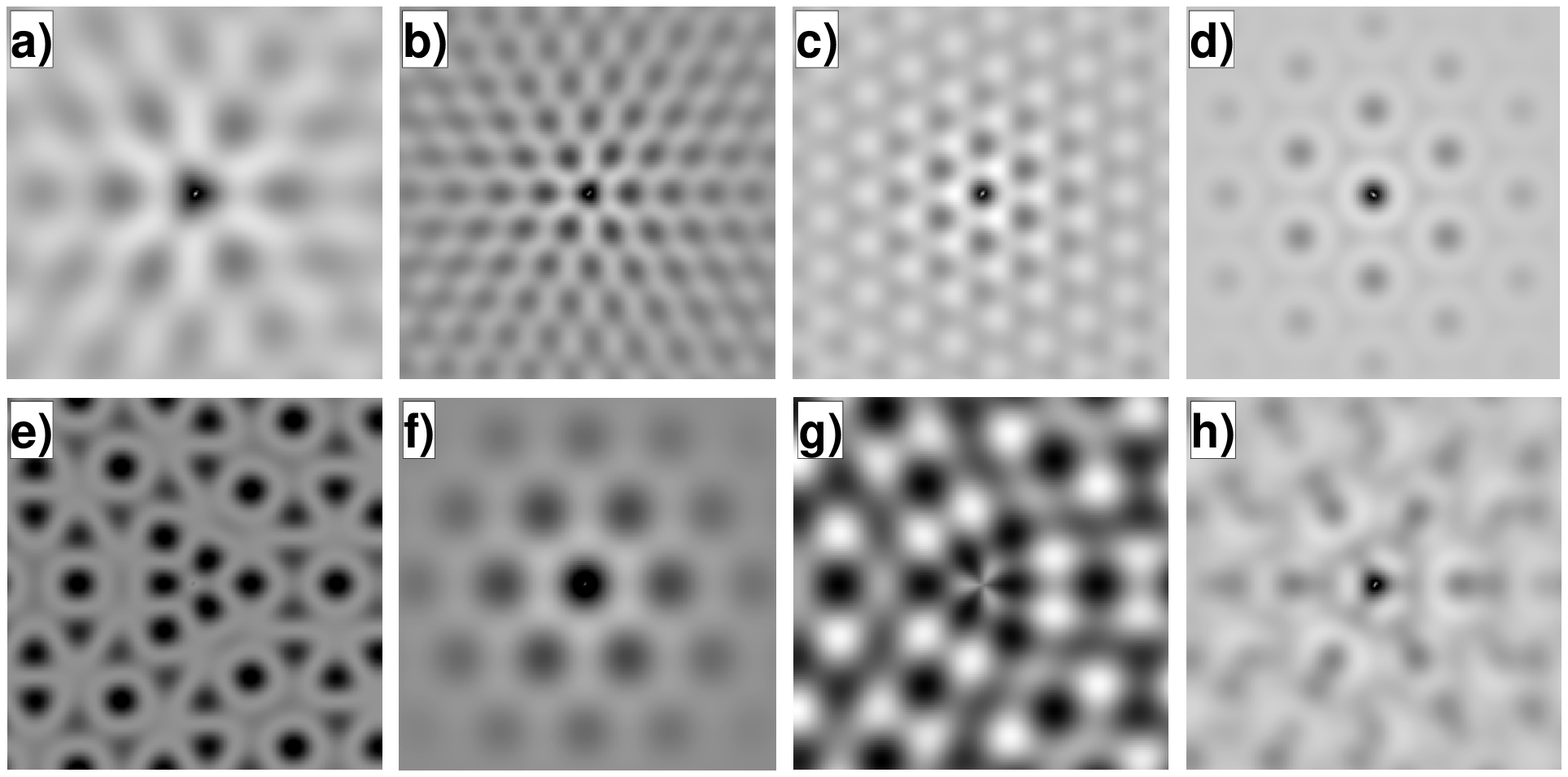}\\
\end{center}
\vspace{0in} \caption{Calculated real-space LDOS modulations in a
monolayer graphene for a point defect placed on top of an A atom.
Figs.~14 a), b), c) depict the contributions of selected scattering
processes to the LDOS ($K \rightarrow K_{1}, K \rightarrow K_{2}$
and $K \rightarrow K_{3}$ respectively) as indicated in
Fig.~5 d); Figs.~14 f), and g) depict the separate contributions of
the A and B sublattices to the $K\rightarrow K_1$ LDOS modulations
in a). Figs.~14 d) and e) depict the A and B sublattice contributions
to the full LDOS, evaluated when all scattering processes are
considered. In h) we depict the sum of the A and B contributions
to the full LDOS, when the weight of the $K\rightarrow K_1$  B
component is increased three times compared to that of the
$K\rightarrow K_1$ A component.}\label{fig8}
\end{figure}

Figure \ref{fig8} displays the calculated real-space modulations
in the LDOS close to the point defect, when only specific
scattering processes have been considered. For simplicity this
calculation has been performed for a monolayer graphene, as for
bilayer graphene it is less clear how much the two bands and the
two layers contribute to the measured LDOS. The intra-nodal
scattering processes are responsible for radially-symmetric
features with small $\overrightarrow{q}$ wavevectors and should
therefore not be responsible for the presence of the three-fold
symmetry.
Fig.~\ref{fig8}.h) depicts the sum of the A
and B contributions to the full LDOS, where in order to increase
the anisotropy, the $K\rightarrow K_1$ B contribution has been
multiplied by three. This may mimic bilayer graphene where, due to
the coupling between the two layers, the sublattices A and B may
not contribute equally to the observed LDOS. A very strong
threefold symmetry is observed in this weighted superposition, and
the resulting image corresponds more closely to the pattern
observed experimentally in figure \ref{fig7}.b). This is consistent
with having a defect at an A site whose dominant effect is the
scattering of sublattice B electrons between two consecutive
valleys ($K \rightarrow K_1$).

More recent experiments presented in \cite{Rutter2} indicate that
in monolayer graphene patterns similar to the ones observed in
\cite{benasimonEPJB} but six-fold symmetry may arise as a result
of rotating sequences of dislocations that close on themselves,
forming grain boundary loops that either conserve the number of
atoms in the hexagonal lattice or accommodate vacancy/interstitial
reconstruction, while leaving no unsatisfied bonds. It would be
interesting to generalize the results presented in \cite{Rutter2}
to bilayer graphene to see if one can obtain similar patterns to
the ones observed in \cite{benasimonEPJB}.

\subsection{Van Hove extension deduced from FT-STS features on
graphene modified by the intercalation of gold clusters.}

\begin{figure}
\begin{center}
\includegraphics[width=15cm]{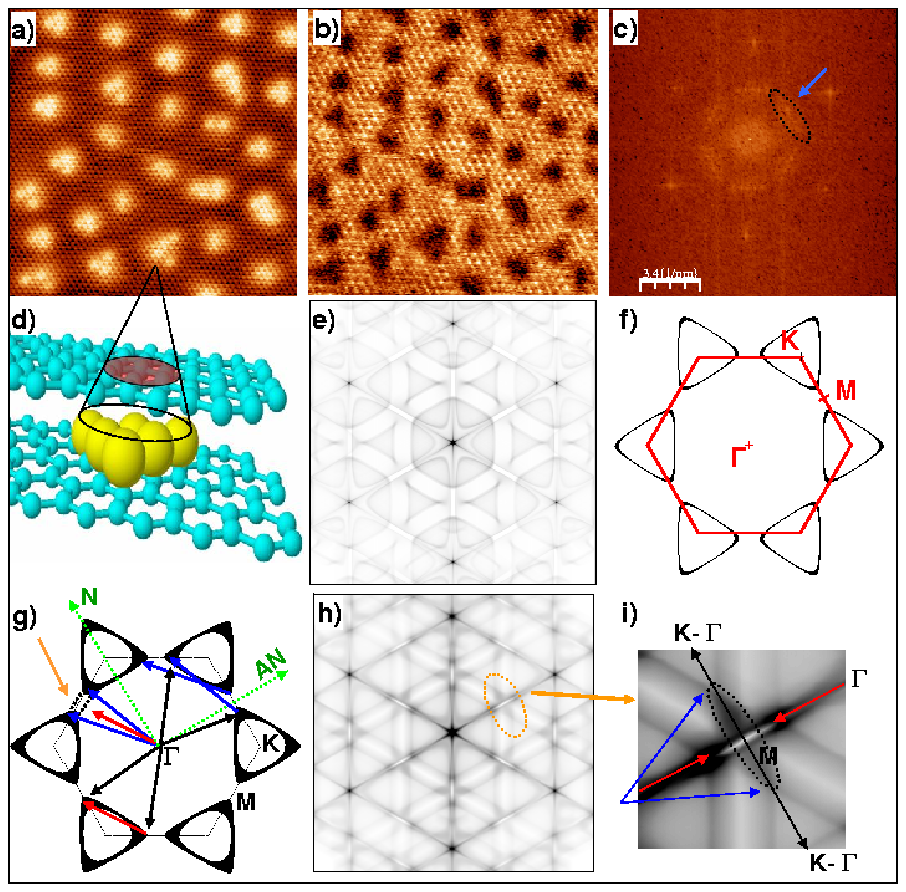}\\
\end{center}
\vspace{0in} \caption{a) Topographic image taken at -1.4V
($13\times13 nm^{2}$) of ML graphene with the intercalated gold
clusters underneath. b) Same region dI/dV map at +0.9V (unoccupied
states). c) FT of b). e) and h) are the JDOS calculation of the
CEC given in f) and g) respectively. i) is a zoom of the ellipitic
feature encircled in h). Blue and red arrows correspond to the
momentum vectors associated to the black vectors. The expected
features in the JDOS calculation that are not observed
experimentally are associated to momentum symbolized by red
arrows. }\label{figintercal}
\end{figure}

The FT-STS technique has recently also been used to interpret the
strong standing waves pattern observed on epitaxial monolayer (ML)
graphene, modified by the intercalation of gold atoms. As
described also above, epitaxial graphene on SiC(0001) consists of
a buffer graphene layer which is covalently bonded with the
substrate, and of a ML graphene weakly connected to the buffer
layer. The monolayer graphene is n-doped with a transfer of
electrons from the substrate. We have discovered that the
deposition of gold atoms under UHV at room temperature, followed
by an annealing cycle, leads to the intercalation of gold in
different forms \cite{PremlalAPL09}. One of them is the
intercalation of aggregates made of 1 to 3 small flat 6-atoms gold
clusters. Figure \ref{figintercal} shows these results. In a) is
shown a topographic image taken at -1V. The clusters appear under
the graphene ML as bright spots. The dI/dV map image of the same
area taken at $+0.9$V shows that the clusters appear as dark
regions. As we probe here the empty states it seems that the
aggregates create an excess of electron as schematized in d).

In between the gold clusters, a standing waves pattern develops,
as revealed by bright p(2x2) protrusions, and the size of the dark
regions decreases as the bias voltage increases
\cite{cranneyEPL10}. These standing waves are associated to
elliptic features in the FT-STS spectra, located around the M
points as shown in c). The size of these elliptic features
increases linearly with the bias voltage. Using the JDOS
calculation we have deduced that these features are associated to
the VHs \cite{cranneyEPL10}. In fact these elliptic features are
observed in the JDOS spectra only when the CEC touch each other at
the M point. This is shown in the calculated JDOS e) and h) for
the CEC f) and g) respectively. In f) the contours do not touch
each other, and no features are observed around M in e) while for
h) as soon as the contours touch each other, an ellipse is
observed.

This ellipse is associated to the k-vectors symbolized as blue
arrows, which connect the apex of two consecutive triangular K
contours (inter-valley scattering). The size of the ellipse is
associated to the filling of the states located at the triangle
apex. This indicates an extension of the VHs. This has been very
recently confirmed by ARPES measurements in the occupied states
\cite{tobepublished}. Two other features in the perpendicular
direction $\Gamma-M-\Gamma$ expected in both JDOS, and even in
T-matrix approximation calculations are not observed
experimentally. These features are associated to the wavevectors
symbolized by the red arrows which should correspond to
intra-valley scattering. As we are at large energies this could
not be due to the pseudospin orientation. The possibility to a
nodal-antinodal dichotomy has been discussed but remains under
debate \cite{cranneyEPL10}. The reasons why the clusters create
such an VHs extension remains non understood. We have recently
shown that a change in the third nearest neighbor hopping energy
in the tight-binding Hamiltonian create strong modifications of
the band structure around the M points and even give rise to new
Dirac points \cite{benasimonPRB2011}, but the experimentally
observed third-nearest neighbor
 coupling seems at present too small to justify important modifications of the band structure for our system.
  More experimental studies and
ab-initio calculations are currently undergoing to clarify the nature
of these standing waves.

\section{General conclusion}
We have shown that simple JDOS calculations and their comparison
with the FT-STS maps provide an accurate determination of the size
and shape of circular, and even non-circular free electron-like
constant energy contours. These contours in turn provide precise
information about the quasiparticle dispersion, and fine details
about fairly complex 2D band structures. On the other hand, we have shown that the T-matrix
approximation becomes absolutely necessary for more complex
systems where more than one quantum number is involved in the
scattering process. 

While in order to apply the T-matrix approximation one needs to have the exact form of the tight-binding Hamiltonian of a system, the JDOS approach does not rely on the exact form of the Hamiltonian, but on the phenomenological form of the equal energy contours. Its advantage is that it can be used even if the Hamiltonian of a system is not known, but the equal energy contours are known from a different experiment such as ARPES; this allows one to have an intuitive reading of the FT features. The results of the T-matrix and the JDOS approximations are in general the same for a symmetrical, quadratic-dispersing system; for example, for the case of Au(111), while not shown here, both the T-matrix and the JDOS approximations yield circles of high intensity in the FT-STS spectra. However, for more complicated systems, these two methods may yield different results, and one notes that when the features predicted by JDOS are not observed experimentally, as it is the case of graphene, the physics of the system is in general more complicated, and a full T-matrix calculation is needed to describe it.

We notice that in general a good agreement
with angle resolved photoemission data has been found and this
indicates that both methods (ARPES and FT-STS) probe the same
quasiparticle excitations of the system. This technique provides
the possibility to measure the band dispersion modification
generated by specific defects of different nature. This allows to
test locally the effect of specific modifications and
functionalization of well-known surfaces which is not possible
with ARPES measurements which requires to prepare large
homogeneous surface in order to obtain reliable data.

\acknowledgments This work is supported by the R\'{e}gion Alsace
and the CNRS, as well as by the ERC Starting Grant NANO-GRAPHENE 256965 . The Agence Nationale de la Recherche supports this
work under the ANR Blanc program, reference
ANR-2010-BLAN-1017-ChimiGraphN, and under the P'NANO program,
reference NANOSIMGRAPHENE. We thank G. Gewinner and F. Gautier for
very useful discussions and J.C. Peruchetti, S. Zabrocki, P.B.
Pillai, M. Narayanan Nair, M.M. De Souza for their contributions
to this work. We thank E. Denys, A. Florentin and A. Le Floch for
the technical support.

\newpage
\end{document}